\documentclass[pra,aps,onecolumn,superscriptaddress,10pt]{quantumarticle}
\usepackage{soul}
\usepackage{tikz}
\usetikzlibrary{shapes.geometric,calc}
\usepackage{relsize}
\usepackage{amsfonts}
\usepackage[sort&compress,numbers,merge]{natbib}
\usepackage{graphicx}
\usepackage{xcolor}
\usepackage{bbm}
\usepackage[bold]{hhtensor}

\usepackage[uselistings]{revquantum}
    \newaffil{IQCAMath}{
        Institute for Quantum Computing,
        Department of Applied Mathematics,
        University of Waterloo,
        Waterloo, ON, Canada
    }
    \newaffil{UQMathPhys}{
        School of Mathematics and Physics, University of Queensland, Brisbane, QLD, Australia
    }
    \newaffil{MITCTP}{
        Center for Theoretical Physics, Massachusetts Institute of Technology, Cambridge, USA
    }

\newcommand{\op}[2]{\ket{#1}\!\bra{#2}}

\newcommand{\dg}{^\dag}
\newcommand{\De}{\widetilde{D}}
\newcommand{\En}{\widetilde{E}}
\newcommand{\Rec}{\mathcal{R}}
\newcommand{\Recb}{\widetilde{\mathcal{R}}}
\newcommand{\Meaz}{\Pi}
\newcommand{\Meazb}{\Pi_\bot }
\newcommand{\LLM}{\logical{\Lambda}_{\logical{\mathrm{M}}}}
\newcommand{\LLP}{\logical{\Lambda}_{\logical{\mathrm{P}}}}
\newrm{No}
\newrm{Un}
\newrm{Co}
\newrm{eff}

    \newcommand{\apxref}[1]{\hyperref[#1]{Appendix~\ref{#1}}}

\colorlet{logical}{cud-orange}
\colorlet{syndrome}{cud-blue}
\colorlet{physical}{gray}
\newcommand{\logical}[1]{{\color{logical} #1}}
\newcommand{\syndrome}[1]{{\color{syndrome} #1}}
\newcommand{\physical}[1]{{\color{physical} #1}}
\newcommand{\Hil}{\ensuremath{\mathcal{H}}}
\newcommand{\HilL}{\ensuremath{\logical{\Hil_L}}}
\newcommand{\HilS}{\ensuremath{{\color{syndrome}{\Hil_S}}}}
\newcommand{\ketl}[1]{\ket{{\color{logical} #1}}}
\newcommand{\kets}[1]{\ket{{\color{syndrome} #1}}}
\newcommand{\etal}{\emph{et al.}}

\theoremstyle{definition}
    \newtheorem{Remark}{Remark}

\tikzset{
    triangle base/.style = {
        regular polygon,
        regular polygon sides=3,
        inner sep=0pt,
        minimum size=0.95em,
        text width=1.2em,
        align=center
    },
    meas/.style = {
        triangle base,
        shape border rotate=-30
    },
    prep/.style = {
        triangle base,
        shape border rotate=30
    },
    gate/.style = {
        rectangle,
        inner sep=0pt,
        text width=1.8em,
        align=center
    },
    two-qubit gate/.style = {
        gate,
        minimum height=3em
    },
    single-qubit gate/.style = {
        gate,
        minimum height=1.2em
    },
    wide single-qubit gate/.style = {
        single-qubit gate,
        text width=2.4em,
        minimum width=2.4em
    },
    physical/.style = {fill=gray!20},
    logical/.style = {fill=logical},
    syndrome/.style = {fill=syndrome, text=white},
    qwire/.style = {line width=1},
    pwire/.style = {qwire, color=physical},
    lwire/.style = {qwire, color=logical},
    swire/.style = {qwire, color=syndrome},
    cwire/.style = {double}
}

\colorlet{link}{cud-bluish-green!35!black!65}

\hypersetup{
  colorlinks   = true, 
  urlcolor     = link, 
  linkcolor    = link, 
  citecolor   = link 
}

\begin{document}

\title{Logical Randomized Benchmarking}

\author{Joshua Combes}
\affilIQCAMath
\affilPI
\affilUQMathPhys

\author{Christopher Granade}
\affilUSydCombined

\author{Christopher Ferrie}
\affilUSydCombined

\author{Steven T. Flammia}
\affilUSydCombined
\affilMITCTP

\date{\today}

\begin{abstract}
    Extrapolating physical error rates to logical error rates requires many
    assumptions and thus can radically under- or overestimate the performance
    of an error correction implementation.  We introduce logical randomized
    benchmarking, a characterization procedure that directly assesses the
    performance of a quantum error correction implementation at the logical
    level, and is motivated by a reduction to the well-studied case of physical randomized benchmarking. 
    We show that our method reliably reports logical performance and
    can estimate the average probability of correctable and uncorrectable errors for a
    given code and physical channel.
\end{abstract}

\maketitle

\section{Introduction}

The accurate estimation of small errors in quantum devices is central to
continuing progress in the field of quantum information.  Several tools have
been developed for this purpose, and one---randomized benchmarking~
\cite{EmerAlicZycz05,KnilLeibReic08,MageGambEmer12}
---has become a \emph{de facto} standard in the field for the estimation of
the quality of quantum gates. To date, nearly all experiments have focused on estimating gate error rates of
physical qubits on a variety of different physical platforms e.g.~\cite{KnilLeibReic08,ChowGambTorn09,RyanLafoLafl09}.  
 Although impressive experiments on characterization of intermediate scale quantum devices have recently been demonstrated~\cite{LuLiTrot15}, those experiements have been forced to make numerous 
oversimplifications of the RB framework to extract relevant information.
As an increasing number of experiments approach the milestone of an encoded or logical qubit~\cite{ChiaLeibScha04,SchiBarrMonz11,ZhanLaflSute12,CorcMageSrin15,KellBareFowl15,RistPoleHuan15,CramKalbRol16,OfekPetrHeer16,TakiCorcMage16}, it is critical to provide useful tools for intermediate-scale characterization.

Here we introduce a method to faithfully assess the performance of logical qubits.
In particular, we develop a protocol to measure the performance
indicators of an error correction implementation that are most important to reasoning
about fault tolerance:
the average probability of both correctable errors and
uncorrectable errors, as well as the average error rate of a gate set.
Our protocol is called  \emph{logical randomized benchmarking} (LRB).
LRB
improves upon current characterization methods by operating directly at the logical, rather than physical, level of
an error correction implementation.  
By performing benchmarking inside this smaller space,
we achieve a drastic reduction in the experimental resources required to characterize an error corrected quantum device.  
Additionally we obtain a more accurate approximation of these performance indicators
with fewer assumptions. 
That is, we
honestly include the effects of imperfect logical gates, crosstalk, and
correlated errors, which are typically ignored.

Our specific protocol is justified by the fact that the proof of correctness reduces to the proof of correctness for the standard case of physical randomized benchmarking. 
This implies additional advantages as well: our protocol remains compatible with the many recent advances
derived from traditional (physical) randomized benchmarking 
\cite{MageGambJohn12,GambCorcMerk12,KimmSilvRyan14,WallBarnEmer14,WallGranHarp15,CrosMageBish16,DugaWallEmer15,HarpFlam16}. 
In addition, this means that the standard and well-known randomized benchmarking assumptions must apply for our analysis to be valid: we require the noise to be time-independent, Markovian, and gate-independent (all at the level of logical Clifford gates).
Unique to the situation of logical qubits, we must also assume that syndrome measurement errors are uncorrelated between rounds of error correction. 
Recently, \citet{ProcRudiYoun17} have demonstrated the importance of the original three preconditions by providing examples where the fidelity can be dramatically overestimated
by naive implementations of RB if gate independence is violated. 
While these assumptions will only ever be approximately satisfied in practice,
randomized benchmarking in general and our protocol in particular can be shown
to be robust to small violations of assumptions such as gate independence \cite{wallman_randomized_2017}.
Thus, using statistical methods that are in turn robust to bounded errors
\cite[supplemental material, section IV]{WiebGranFerr14}, our protocol provides useful estimates in cases of experimental interest.

In principle the existing
characterization and benchmarking toolkits can be applied directly to the
physical qubits that comprise a logical
qubit~\cite{MageGambJohn12,GambCorcMerk12,KimmSilvRyan14,WallBarnEmer14,WallGranHarp15,CrosMageBish16,DugaWallEmer15,HarpFlam16}.
In practice, however, there are formidable challenges associated with this approach.
Using randomized benchmarking of physical gates to compare to fault-tolerance thresholds is 
notoriously difficult to do honestly
\cite{KuenLongDohe16,Puzz14,WallGranHarp15,SandWallSand16,GutiBrow15,DarmPoul16}. 
In multi-qubit systems additional imperfections, such as crosstalk and spatially
or temporally correlated errors, can make characterization still harder. Indeed, it
need not be the case that a single-qubit error rate is representative of an
error rate in a larger system, and these extrapolated error rates can even be off by many orders of magnitude in some cases~\cite{Poulin2017Coogee}. 
In fact, we will show later that characterization 
methods that operate at the physical layer can either systematically under- 
\emph{or} overestimate the error rates of logical gates.


Our work is inspired by and related to many other works. We briefly comment on the relationship to some of those works.
\citet{AshiBargKnil00} has previously defined the probability for an uncorrectable error. 
\citet{Scott2005} used the error probabilities to compare different quantum error correcting codes of the same minimum distance. 
Neither of these papers gave a procedure to measure the error probabilities. 
In the context of calculating thresholds for concatenated codes,
\citet{RahnDoheMabu02} (and later work by~\citet{HuanYouWu15}) assumed a particular physical error channel and successively mapped this to logical channels with concatenated codes. 
Our proposal inherits the intuition built by up by Rahn~\etal. 
One of the key differences in our approaches is that we assume nothing about the physical channel. 
Instead, we advocate measuring properties of logical channels directly. 

The remainder of the paper is structured as follows. 
We start by defining the average
probability of a correctable error and uncorrectable error for a quantum error
correcting code. Then we relate these probabilities to the logical fidelity
of a channel, and give a protocol to measure the logical fidelity. This
protocol is inherently sensitive to state preparation and measurement (SPAM)
errors. To solve the sensitivity to SPAM errors we introduce the LRB protocol.
Then we finish with a discussions of the advantages of LRB over physical RB.

Throughout the paper, we use the notational conventions that unitary operators
are written as roman letters ($U$), with corresponding quantum channels written in calligraphic
script (so that $\mathcal U(\rho) \defeq U \rho U^\dagger$), and with encoded operators and channels denoted by bars
($\overline U$, $\overline{\mathcal U}$). Similarly, imperfect or noisy channels
corresponding to $U$ or $\mathcal U$ are written using tildes ($\widetilde{\mathcal U}$).
We emphasize the important special case of noise with respect to the logical or
physical identity channels by denoting these channels as $\Lambda$ or
$\overline \Lambda$, respectively. Finally, we will 
denote expectation values of random variables by $\expect$, with subscripts being used to indicate
which variable an expectation is taken over if it is not otherwise clear.

\section{Errors and Logical Fidelity}

We seek a protocol that measures how well a given quantum error correcting code
protects against an error channel $\overline{\Lambda}$. Consider a known state $\ket{\psi}$ and its
encoding $\ket{\overline{\psi}} \defeq E (\ket{\psi} \otimes \ket{0})$ into the relevant codewords using an encoding map $E$ and an ancilla initially prepared in the state $\ket{0}$. 
The encoded state is then subject to the noise channel, the recovery operation $\mathcal{R}$,
and finally the decoder $D = E^\dagger$.
We can measure how well our code protects against the noise 
$\overline{\Lambda}$ by measuring
$\overline{\Meaz} \defeq \op{\overline{\psi}}{\overline{\psi}}$
versus $\overline{\Meaz}_{\perp} = \id - \op{\overline{\psi}}{\overline{\psi}}$.
We adopt the notation that
\begin{align}
    \Pr(\overline{\Meaz} | \psi, \mathcal R, \overline{\Lambda}) 
    \defeq \Tr\left[
        \op{\overline{\psi}}{\overline{\psi}} \mathcal{R} \overline{\Lambda}(\op{\overline{\psi}}{\overline{\psi}})
    \right].
\end{align}
Using this notation, we can now define the probability that no errors
have occurred in terms of the probability that we corrected an error
using an identity channel as the recovery map. That is,
\begin{align}\label{eq:no_err}
    \Pr(\No | \psi, \overline{\Lambda})  
    & \defeq \Pr(\overline{\Meaz} | \psi, \mathcal R=I, \overline{\Lambda}).
\end{align}
When a non-trivial recovery operation is performed,
we can then break down the probability of correctly measuring
the state we prepared into the probabilities that no error occurred,
and that we corrected an error that did occur.
Formally, we write that the probability of an uncorrectable error is~\cite{AshiBargKnil00,Scott2005}
\begin{align}\label{eq:un_err}
    \Pr(\Un| \psi, \mathcal{R}, \overline{\Lambda}) \defeq 1 -
            \Pr(\overline{\Meaz} | \psi, \mathcal R, \overline{\Lambda}).
\end{align}
We complete our definition of the relevant code properties
by taking the probability of a correctable error to be
\begin{align}\label{eq:co_err}
    \Pr(\Co | \psi, \mathcal{R}, \overline{\Lambda}) \defeq
      1-  
          \Pr(\Un| \psi, \mathcal{R}, \overline{\Lambda}) 
      -\Pr(\No | \psi, \overline{\Lambda}).
\end{align}

In practice, however, learning code properties by directly estimating these parameters is not feasible,
as it requires perfect state preparation and measurement (SPAM), and is sensitive to
the specific choice of input state $\ket{\psi}$. To solve this, we will show how applying
randomized benchmarking to encoded gates will allow us to estimate the
\emph{average gate fidelity}
\begin{align}\label{eq:LogFid}
    F_L( \mathcal{R} \overline{\Lambda}) \defeq
    F(\mathcal{R} \overline{\Lambda}, \id) = 
        \int \dd\psi \braket{
            \overline{\psi} |
            \mathcal{R} \overline{\Lambda}\left(
                \op{\overline{\psi}}{\overline{\psi}}
            \right) |
            \overline{\psi}
        },
\end{align}
where $\dd\psi$ is the Haar measure
and $F(\mathcal{R} \overline{\Lambda}, \id)$ is the fidelity between the
channel $\mathcal{R} \overline{\Lambda}$ and the identity.
Since 
\begin{subequations}
\begin{align}
F(\mathcal{R} \overline{\Lambda}, \id) &=
\expect_{\psi}[\Pr(\No | \psi, \overline{\Lambda}) + \Pr(\Co | \psi,\mathcal{R}, \overline{\Lambda})] \,\,\textrm{and}\\
 F( \overline{\Lambda}, \id) &= 
\expect_{\psi}[\Pr(\No | \psi, \overline{\Lambda}) ],
\end{align}
\end{subequations}
 our protocol
directly enables estimating average performance indicators of quantum
error correcting codes and their associated recovery operators.

To use the average gate fidelity in the context of quantum error correction, we consider
an encoded logical unitary $\overline{U}$ for a unitary $U$,
and will consider the case in which the code under consideration encodes
$k$ logical qubits into $n$ physical qubits. For an arbitrary $k$-qubit
state $\rho$, we will thus write that the corresponding $n$-qubit
encoded state $\overline{\rho}$ is given by $E (\rho \otimes \phi) E\dg$,
where $\phi$ is a fiducial state for an ancilla register. That is,
the encoder $E$ effectively changes bases such that such that
the code space $\Hil_{\mathrm{C}}$ factorizes as
the tensor product $\mathcal{H}_L \otimes \mathcal{H}_S$ of a logical space $\mathcal{H}_L$
and a syndrome space $\mathcal{H}_S$.
For this reason, we can consider $E$ and $D$ to be perfect; they act to
\emph{define} the relevant spaces, rather than denote physical actions
taken during an experiment.

The encoded unitary $\overline{U}$ is related to $U$ by
conjugation by the decoder,
$\overline{U} \defeq E U D = D\dg U D$.
We model an imperfect implementation of $\overline{U}$ as the
ideal encoded unitary followed by a \emph{physical} error channel acting on the encoded degrees of freedom:
$\widetilde{\mathcal{U}}=\overline{\Lambda}\,\overline{\mathcal{U}}$.
We now use the encoder $E$ as a mathematical tool to reason about the logical channel,
rather than as a description of a physical operation.
That is, the encoding and decoding operations \emph{need only exist abstractly},
such that our protocol does \emph{not} require experimentally implementing
encoding or decoding, except for at the initial preparation and
final measurement steps.
The channels $\overline{\mathcal U}$ and
$\widetilde{\mathcal U}$ can be represented as the quantum circuits 

\def\offseta{6.5}
\def\offsetap{0.5}
\def\offsetb{1} 
\def\offsetc{3.7} 
\begin{center}
    \tikz[baseline=-1.3ex]{ 
        \begin{scope}[every node/.style={scale=0.85}, scale=0.85]
            \draw[pwire] (-\offseta-0.5,  0.25) -- (-\offseta+0.5, 0.25);
            \draw[pwire] (-\offseta-0.5, -0.25) -- (-\offseta+0.5, -0.25);
            \draw[pwire] (-\offseta-0.5, -0.75) -- (-\offseta+0.5, -0.75);
            \node[gate, physical, minimum height=4em] at (-\offseta+0, -0.25) {$\overline{U}$};
            \node at (-\offseta+1, -0.25) {$=$};
            \draw[pwire] (-\offsetap-2, 0.25) -- (-\offsetap-1.5, 0.25);
            \draw[pwire] (-\offsetap-2, -0.25) -- (-\offsetap-1.5, -0.25);
            \draw[pwire] (-\offsetap-2, -0.75) -- (-\offsetap-1.5, -0.75);
            \draw[pwire] (-\offsetap-4, 0.25) -- (-\offsetap-2, 0.25);
            \draw[pwire] (-\offsetap-4, -0.25) -- (-\offsetap-2, -0.25);
            \draw[pwire] (-\offsetap-4, -0.75) -- (-\offsetap-2, -0.75);
            \draw[pwire] (-\offsetap-4.5, 0.25) -- (-\offsetap-4, 0.25);
            \draw[pwire] (-\offsetap-4.5, -0.25) -- (-\offsetap-4, -0.25);
            \draw[pwire] (-\offsetap-4.5, -0.75) -- (-\offsetap-4, -0.75);
            \node[gate, physical, minimum height=4em] at (-\offsetap-2, -0.25) {$D$};
            \node[gate, physical, minimum height=1.5em] at (-\offsetap-3, 0.25) {$U$};
            \node[gate, physical, minimum height=4em] at (-\offsetap-4, -0.25) {$E$};

            \node[] at (-0.75, -0.25) {and};

            \draw[pwire] (\offsetb-0.5,  0.25) -- (\offsetb+0.5, 0.25);
            \draw[pwire] (\offsetb-0.5, -0.25) -- (\offsetb+0.5, -0.25);
            \draw[pwire] (\offsetb-0.5, -0.75) -- (\offsetb+0.5, -0.75);
            \node[gate, physical, minimum height=4em] at (\offsetb+0, -0.25) {$\widetilde{\mathcal U}$};
            \node at (\offsetb+1.125, -0.25) {$=$};       
            \draw[pwire] (-1+\offsetc, 0.25) -- (1+\offsetc, 0.25);
            \draw[pwire] (-1+\offsetc, -0.25) -- (1+\offsetc, -0.25);
            \draw[pwire] (-1+\offsetc, -0.75) -- (1+\offsetc, -0.75);
            \node[gate, physical, minimum height=4em] at (0.5+\offsetc, -0.25) {$\overline{ U }$};
            \node[gate, physical, minimum height=4em] at (-0.5+\offsetc, -0.25) {$\overline{\Lambda}$};
        \end{scope}
    }\ .
\end{center}

\noindent
Here, and throughout, we adopt a number of conventions: 
circuit diagrams are drawn in the Kitaev convention with time $t$ increasing to the left, 
the encoder and decoder leave the state of interest on the top wire, 
and finally the lower two wires are the syndrome wires and represent as many wires as are necessary to carry this information. 
In the first circuit we have assumed a particular convention for the encoded
unitary $\overline{U}$ and the syndrome space $\mathcal H_S$, namely that we choose representatives
for each logical unitary that preserve our choice of basis for $\mathcal H_S$, see \autoref{rem:ed_conven} in the Appendix for more details.

We now present a protocol to measure the logical fidelity as defined in \autoref{eq:LogFid}.
Here and throughout we only consider direct measurements of syndromes, as it is straightforward to reduce the
logical channel of an ancilla-coupled syndrome readout to the direct case; we show this in
detail in \apxref{apx:ancilla_coupled}.
Given this convention, one possible approach to measure the logical fidelity would
then be to prepare random states $\ket{\psi}$, evolve according
to the unknown gates, and then measure whether that state is
preserved. In circuit form, we draw this as
\begin{align}
    \label{cir:logfid}
    \tikz[baseline=2ex]{ 
        \begin{scope}[every node/.style={scale=0.86}, scale=0.86]
            \node[] at (1, 1.0) {$\ket{\psi}$};
            \node[] at (1, 0.5) {$\ket{0}$};
            \node[] at (1, 0.0) {$\ket{0}$};
            \draw[pwire] (0.5, 1.0) -- (-5, 1.0);
            \draw[pwire] (0.5, 0.5) -- (-5, 0.5);
            \draw[pwire] (0.5,  0.0) -- (-5, 0.0);
            \node[gate, physical, minimum height=4em] at (0, 0.5) {$\mathcal\En$};
            \node[gate, physical, minimum height=4em] at (-1, 0.5) {$\widetilde {\mathcal U}$};
            \node[gate, physical, minimum height=4em] at (-2, 0.5) {$\Recb$};
            \node[gate, physical, minimum height=4em] at (-3, 0.5) {$\mathcal\De$};
            \node[gate, physical, minimum height=1.3em,minimum width=2em] at (-4, 1) {$U^{-1}$};
            \node[scale=0.7] at (-5.5, 1) {$\Meaz$ vs $\Meazb$};
            \node[meas, physical, scale=0.8] at (-5, 0.25) {${\Tr}$};
        \end{scope}
    },
\end{align}
where we have implicitly defined the \emph{logical} channel as that which
maps the input state on the top wire to the output state also on the top wire. 
As stated before, the noisy operations are denoted by tildes, i.e.\ encoding $\mathcal\En$, decoding 
$\mathcal \De $, logical unitary $\widetilde {\mathcal U}$, and recovery $\Recb$.
Explicitly, the logical channel is defined as
\begin{align}
	\Lambda_L(\rho_L) \defeq \Tr_{\rm synd}\big [\mathcal{U}^{-1} \mathcal\De  \Recb  \widetilde{\mathcal{U}}  \mathcal \En  (\rho_L\otimes \op{0}{0}_{\rm synd}) \big] 
\end{align}
for an input logical state $\rho_L$.
To estimate the logical fidelity we repeat this experiment many times
with different input states $\psi$. If we choose $\psi$ from a spherical
2-design such as the stabilizer states, then this procedure provides a Monte Carlo approximation to
the integral that defines the logical channel fidelity of \autoref{eq:LogFid}.

Unfortunately, this approach does not separate imperfections in the encoder and decoder
from imperfections in the gate of interest. Encoding and decoding, as we argued earlier,
should be thought of as a change of basis, rather than as actual gates.
Thus, to avoid conflating the logical fidelity of interest with state preparation and measurement (SPAM) errors,
and to separate the error from the initial encoding and decoding errors,
we require a more robust approach.
To do this, 
we need to generalize this protocol to repeated rounds of error correction.
Our method is a generalization of randomized 
benchmarking~\cite{EmerAlicZycz05,KnilLeibReic08,MageGambEmer12} (RB), 
as RB has been demonstrated to provide useful fidelity estimates even in the presence of
strong SPAM errors.

\section{Logical Randomized Benchmarking} 

\begin{figure*}[ht]
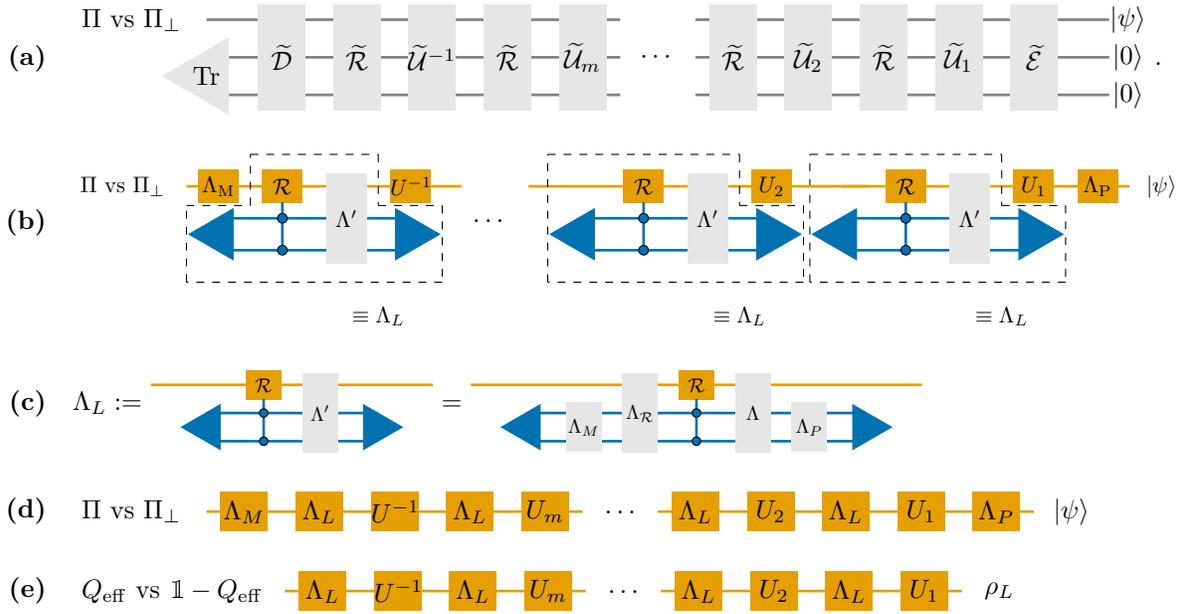

    \begin{centering}
        \begin{align*}
            \textbf{(a)} \quad
            & \tikz[baseline=-1em]{
                \draw[pwire] (-1.5, 0.25) -- (4, 0.25);
                \draw[pwire] (-1.5, -0.25) -- (4, -0.25);
                \draw[pwire] (-1.5, -0.75) -- (4, -0.75);
                \draw[pwire] (-8, 0.25) -- (-2.5, 0.25);
                \draw[pwire] (-8, -0.25) -- (-2.5, -0.25);
                \draw[pwire] (-8, -0.75) -- (-2.5, -0.75);
                \node[gate, physical, minimum height=4em] at (3, -0.25) {$\mathcal\En$};
                \node[gate, physical, minimum height=4em] at (2, -0.25) {$\widetilde{\mathcal U}_1$};
                \node[gate, physical, minimum height=4em] at (1, -0.25) {$\Recb$};
                \node[gate, physical, minimum height=4em] at (0, -0.25) {$\widetilde{\mathcal U}_2$};
                \node[gate, physical, minimum height=4em] at (-1, -0.25) {$\Recb$};
                \node[] at (-2, -0.25) {$\cdots$};
                \node[gate, physical, minimum height=4em] at (-3, -0.25) {$\widetilde{\mathcal U}_m$};
                \node[gate, physical, minimum height=4em] at (-4, -0.25) {$\Recb$};
                \node[gate, physical, minimum height=4em] at (-5, -0.25) {$\widetilde{ \mathcal U}^{-1}$};
                \node[gate, physical, minimum height=4em] at (-6, -0.25) {$\Recb$};
                \node[gate, physical, minimum height=4em] at (-7, -0.25) {$\mathcal\De$};
                \node[] at (4.25, 0.25) {$\ket{\psi}$};
                \node[] at (4.25, -0.25) {$\ket{0}$};
                \node[] at (4.25, -0.75) {$\ket{0}$};
                \node[] at (-9, 0.25) {$\Meaz$ vs $\Meazb$};
                \node[meas, physical] at (-8, -0.5) {${\Tr}$};
            }. \\[1em]
            \textbf{(b)} \quad & \tikz[baseline=-1em]{
                \foreach \dx in {1} {
                    \begin{scope}[every node/.style={scale=0.85}, scale=0.85, shift={(18 -6* \dx , 0)}]
                        \node[] at (-9, 0.25) {$\ket{\psi}$};
                        \node[prep, syndrome, scale=0.8] at (-11, -0.5) {};
                        \draw[lwire] (-14, 0.25) -- (-9.5, 0.25);
                        \draw[swire] (-13.8, -0.25) -- (-11.1, -0.25);
                        \draw[swire] (-13.8, -0.75) -- (-11.1, -0.75);
                        \node[gate, logical, minimum height=1.5em] at (-10, 0.25) {$\Lambda_{\rm P}$};
                        \node[gate, logical, minimum height=1.5em] at (-11, 0.25) {$U_{\dx}$};
                        \node[gate, physical, minimum height=4em] at (-12, -0.25) {$\Lambda'$};
                        \node[gate, logical, minimum height=1.5em] at (-13, 0.25) {$\mathcal R$};
                        \draw[swire]  (-13, 0) -- (-13, -0.75);
                        \draw[fill=syndrome] (-13, -0.25)  circle (0.075);
                        \draw[fill=syndrome] (-13, -0.75)  circle (0.075);
                        \node[meas, syndrome, scale=0.8] at (-14, -0.5) {};
                        \draw[dashed]       (-14.5,0.75) -- (-11.5,0.75)  ;
                        \draw[dashed]       (-11.5,0.75) -- (-11.5,-0.05)   ;
                        \draw[dashed]       (-11.5,-0.05) -- (-10.5, -0.05)  ;
                        \draw[dashed]       (-10.5, -0.05) -- (-10.5, -1.25)  ;
                        \draw[dashed]       (-10.5, -1.25) -- (-14.5, -1.25) ;
                        \node[below] at (-11.5,-1.5) { $\equiv\Lambda_L$};
                        \draw[dashed]       (-14.5, -1.25)--(-14.5,0.75);
                    \end{scope}
                }
                \foreach \dx in {2} {
                    \begin{scope}[every node/.style={scale=0.85}, scale=0.85, shift={(19.9 -6* \dx , 0)}]
                        \node[prep, syndrome, scale=0.8] at (-11, -0.5) {};
                        \draw[lwire] (-14.8, 0.25) -- (-9.5, 0.25);
                        \draw[swire] (-13.8, -0.25) -- (-11.1, -0.25);
                        \draw[swire] (-13.8, -0.75) -- (-11.1, -0.75);
                        \node[gate, logical, minimum height=1.5em] at (-11, 0.25) {$U_{\dx}$};
                        \node[gate, physical, minimum height=4em] at (-12, -0.25) {$\Lambda'$};
                        \node[gate, logical, minimum height=1.5em] at (-13, 0.25) {$\mathcal R$};
                        \draw[swire]  (-13, 0) -- (-13, -0.75);
                        \draw[fill=syndrome] (-13, -0.25)  circle (0.075);
                        \draw[fill=syndrome] (-13, -0.75)  circle (0.075);
                        \node[meas, syndrome, scale=0.8] at (-14, -0.5) {};
                        \draw[dashed]       (-14.5,0.75) -- (-11.5,0.75)  ;
                        \draw[dashed]       (-11.5,0.75) -- (-11.5,-0.05)   ;
                        \draw[dashed]       (-11.5,-0.05) -- (-10.5, -0.05)  ;
                        \draw[dashed]       (-10.5, -0.05) -- (-10.5, -1.25)  ;
                        \draw[dashed]       (-10.5, -1.25) -- (-14.5, -1.25) ;
                        \node[below] at (-11.5,-1.5) { $\equiv\Lambda_L$};
                        \draw[dashed]       (-14.5, -1.25)--(-14.5,0.75);
                    \end{scope}
                }
                \node at (-6.4, -0.25) {$\ldots$};
                \begin{scope}[every node/.style={scale=0.85}, scale=0.85, shift={(2.25 , 0)}]
                    \draw[lwire] (-14.5, 0.25) -- (-10.2, 0.25);
                    \draw[swire] (-14, -0.25) -- (-11, -0.25);
                    \draw[swire] (-14, -0.75) -- (-11, -0.75);
                    \node[prep, syndrome, scale=0.8] at (-11, -0.5) {};
                    \node[gate, logical, minimum height=1.5em] at (-11, 0.25) {$U^{-1}$};
                    \node[gate, physical, minimum height=4em] at (-12, -0.25) {$\Lambda'$};
                    \node[gate, logical, minimum height=1.5em] at (-13, 0.25) {$\mathcal R$};
                    \draw[swire]  (-13, 0) -- (-13, -0.75);
                    \draw[fill=syndrome] (-13, -0.25)  circle (0.075);
                    \draw[fill=syndrome] (-13, -0.75)  circle (0.075);
                    \node[gate, logical, minimum height=1.5em] at (-14, 0.25) {$\Lambda_{\rm M}$};
                    \draw[dashed]       (-13.5,0.75) -- (-11.5,0.75)  ;
                    \draw[dashed]       (-11.5,0.75) -- (-11.5,-0.05)   ;
                    \draw[dashed]       (-11.5,-0.05) -- (-10.5, -0.05)  ;
                    \draw[dashed]       (-10.5, -0.05) -- (-10.5, -1.25)  ;
                    \draw[dashed]       (-10.5, -1.25) -- (-14.5, -1.25) ;
                    \node[below] at (-11.5,-1.5) { $\equiv\Lambda_L$};
                    \draw[dashed]       (-14.5, -1.25)--(-14.5,-0.05);
                    \draw[dashed]       (-14.5,-0.05)-- (-13.5,-0.05);
                    \draw[dashed]       (-13.5,-0.05) -- (-13.5,0.75);
                    \node[] at (-15.5, 0.25) {$\Meaz$ vs $\Meazb$};
                    \node[meas, syndrome, scale=0.8] at (-14, -0.5) {};
                \end{scope}
            } \\[1em]
            \textbf{(c)} \quad & {\Lambda_L}\defeq  \tikz[baseline=-1.0ex]{
            \begin{scope}[every node/.style={scale=0.75}, scale=0.75, shift={(0 , 0)}]
                \draw[lwire] (-15, 0.25) -- (-10, 0.25);
                \draw[swire] (-14, -0.25) -- (-11, -0.25);
                \draw[swire] (-14, -0.75) -- (-11, -0.75);
                \node[prep, syndrome, scale=0.8] at (-11, -0.5) {};
                \node[gate, physical, minimum height=4em] at (-12, -0.25) {$\Lambda'$};
                \node[gate, logical, minimum height=1.5em] at (-13, 0.25) {$\mathcal R$};
                \draw[swire]  (-13, 0) -- (-13, -0.75);
                \draw[fill=syndrome] (-13, -0.25)  circle (0.075);
                \draw[fill=syndrome] (-13, -0.75)  circle (0.075);
                \node[meas, syndrome, scale=0.8] at (-14, -0.5) {};
            \end{scope}
            }
            = \tikz[baseline=-1.0ex]{
            \begin{scope}[every node/.style={scale=0.75}, scale=0.75, shift={(0 , 0)}]
                \draw[lwire] (-18, 0.25) -- (-10, 0.25);
                \draw[swire] (-17, -0.25) -- (-11, -0.25);
                \draw[swire] (-17, -0.75) -- (-11, -0.75);
                \node[prep, syndrome, scale=0.8] at (-11, -0.5) {};
                \node[gate, physical, minimum height=2.5em] at (-12, -0.5) {$\Lambda_{P}$};
                \node[gate, physical, minimum height=4em] at (-13, -0.25) {$\Lambda$};
                \node[gate, logical, minimum height=1.5em] at (-14, 0.25) {$\mathcal R$};
                \draw[swire]  (-14, 0) -- (-14, -0.75);
                \draw[fill=syndrome] (-14, -0.25)  circle (0.075);
                \draw[fill=syndrome] (-14, -0.75)  circle (0.075);
                \node[gate, physical, minimum height=4em] at (-15, -0.25) {$\Lambda_{\mathcal R}$};
                \node[gate, physical, minimum height=2.5em] at (-16, -0.5) {$\Lambda_{M}$};
                \node[meas, syndrome, scale=0.8] at (-17, -0.5) {};
            \end{scope}
            } \\[1em]
            \textbf{(d)} \quad & \tikz[baseline=0.5em]{
                \draw[lwire] (-16.5, 0.25) -- (-11.5, 0.25);
                \draw[lwire] (-10.5, 0.25) -- (-5.5, 0.25);
                \node[] at (-5, 0.25) { ${\ket{\psi}}$};
                \node[gate, logical, minimum height=1.5em] at (-6, 0.25) {$\Lambda_P$};
                \node[gate, logical, minimum height=1.5em] at (-7, 0.25) {$U_1$};
                \node[gate, logical, minimum height=1.5em] at (-8, 0.25) {$\Lambda_L$};
                \node[gate, logical, minimum height=1.5em] at (-9, 0.25) {$U_2$};
                \node[gate, logical, minimum height=1.5em] at (-10, 0.25) {$\Lambda_L$};
                \node at (-11, 0.25) {$\ldots$};
                \node[gate, logical, minimum height=1.5em] at (-12, 0.25) {$U_m$};
                \node[gate, logical, minimum height=1.5em] at (-13, 0.25) {$\Lambda_L$};
                \node[gate, logical, minimum height=1.5em] at (-14, 0.25) {$U^{-1}$};
                \node[gate, logical, minimum height=1.5em] at (-15, 0.25) {$\Lambda_L$};
                \node[gate, logical, minimum height=1.5em] at (-16, 0.25) {$\Lambda_M$};
                \node[] at (-17.5, 0.25) {$\Meaz$ vs $\Meazb$};
            } \\[1em]
            \textbf{(e)} \quad & \tikz[baseline=0.5em]{
                \draw[lwire] (-15.5, 0.25) -- (-11.5, 0.25);
                \draw[lwire] (-10.5, 0.25) -- (-6.5, 0.25);
                \node[] at (-6, 0.25) { ${\rho_L}$};
                \node[] at (-4.88, 0.25) { \phantom{q}};
                \node[gate, logical, minimum height=1.5em] at (-7, 0.25) {$U_1$};
                \node[gate, logical, minimum height=1.5em] at (-8, 0.25) {$\Lambda_L$};
                \node[gate, logical, minimum height=1.5em] at (-9, 0.25) {$U_2$};
                \node[gate, logical, minimum height=1.5em] at (-10, 0.25) {$\Lambda_L$};
                \node at (-11, 0.25) {$\ldots$};
                \node[gate, logical, minimum height=1.5em] at (-12, 0.25) {$U_m$};
                \node[gate, logical, minimum height=1.5em] at (-13, 0.25) {$\Lambda_L$};
                \node[gate, logical, minimum height=1.5em] at (-14, 0.25) {$U^{-1}$};
                \node[gate, logical, minimum height=1.5em] at (-15, 0.25) {$\Lambda_L$};
                \node[] at (-17, 0.25) {${Q_\eff}$ vs ${\id}-{Q_\eff}$};
            }     
        \end{align*}
    \end{centering}
    \caption{
        \label{fig:protocol}
        A summary of the derivation of logical randomized benchmarking
        protocol from a physical circuit.  See the Appendix for the complete
        derivation.  The derivation proceeds by reducing the physical circuit
        to a logical circuit that looks like Magesan \etal's
        \cite{MageGambEmer12,MageGambEmer11} zeroth-order RB.
        \textbf{(a)} 
        An example physical circuit for a sequence length $m$, following step
        (i) of the experimental protocol described in the main text.
        \textbf{(b)} 
        The circuit of (a), decomposed into logical (\logical{yellow}) and syndrome
        spaces (\syndrome{blue}) with the definitions of the noisy implementations
        $\tilde{U}$, $\tilde{R}$, $\tilde{E}$ and $\tilde{D}$ applied and with
        the unit cell identified. Here $\Lambda'$ is the composition of several
        channels whose definition is given in the Appendix above
        \autoref{eq:lambdaprime}. It is important to our protocol that the
        syndromes are refreshed during rounds of error correction.
        \textbf{(c)} 
        The logical channel $\Lambda_L$, defined in terms of additional noisy
        channels $\Lambda_{\mathcal{R}}$, $\Lambda_M$, and $\Lambda_P$ for the
        syndrome recovery, measurement, and preparation that are defined in
        the Appendix.        
        \textbf{(d)} 
        The circuit of (b), with the syndrome degrees of freedom traced
        out to obtain the logical circuit.        
        \textbf{(e)} 
        The circuit of (d), written in the form of the Magesan \etal~model
        with $\rho_L = \Lambda_P(\op{\psi}{\psi})$ and $Q_\eff =
        \Lambda_M^{*}(\Meaz)$, where $\Lambda_M^*$ is the dual channel
        of $\Lambda_M$.  Notice that the initial encoding is noisy and
        the final decoding is also noisy, and that this noise gets absorbed into
        state preparation and measurement errors.
    }
\end{figure*}

In \autoref{fig:protocol}, we provide a visual summary of the full derivation of our logical randomized benchmarking protocol, provided in \apxref{apx:LRB_derivation}. Our proof is by reduction from our model to the zeroth order RB of Magesan \etal \cite{MageGambEmer12}. 
This is the key feature of our protocol, and distinguishes it from any of several other ``obvious'' ways that one might perform logical error estimation.
Although these other methods might be physically reasonable, our method allows us to build our results on the solid theoretical foundations of standard RB. 
We do not claim that our method is unique or ``best'', but only that it provides a rigorous and justified way to reduce to this most well-studied case. 
We also note that the initial preparation and the final measurement arise differently in our protocol than in traditional RB, but yield the same effective model.

Logical randomized benchmarking consists of enacting sequences of random logical Clifford gates of length $m$, where each sequence is followed by a single gate inverting that sequence. Each sequence of length $m$ is repeated $T$ times, and for each distinct length, we choose $L$ different sequences.  For each $m,t \in \{M,T\}$, the protocol for logical randomized benchmarking is as follows:
\begin{enumerate}
	\item For each $i \in \{1, \dots, L\}$, perform the following:
	\begin{enumerate}
		\item Choose a sequence $s_i = \{U_{i_1}, U_{i_2}, \dots, U_{i_m}\}$ of Clifford operations.
		\item Perform the following $T$ times:
		\begin{enumerate}
			\item Prepare the state $\rho$ and encode into $\overline{\rho}$.
			\item Apply each encoded Clifford operation $\overline{U} \in s_i$, following each with a round of error correction (with or without recovery) 	and a refresh of the ancilla register.
			\item Apply the Clifford operation $\overline{U}^{-1} = \left(\prod_{\overline{U}\in s_i} \overline{U}^\dagger \right)$.
			\item Decode and measure the final state against the observable $Q$.
		\end{enumerate}
		\item Record the average observed value of $Q$ as $q_i$.
	\end{enumerate}
	\item Record the average of $q_i$ over sequences as $\overline{q}(m)$.
\end{enumerate}
Then, following the standard RB argument, the logical sequence fidelity can be determined from
the expectation of $\overline{q}(m)$ over all random
choices and outcomes, which is given by
\begin{align}\label{eq:decay}
  F_L(m)\defeq\expect[\overline{q}(m)] = A_L p_L^m + B_L, 
\end{align}
for constants $A_L$ and $B_L$ defined in the Appendix that are sensitive to \emph{logical} SPAM errors, the decay constant $p_L$ is defined by
\begin{align}
         {p_L} & \defeq \frac{
             (\dim \mathcal H _L) F({\Lambda_L}, {\id}) - 1
         }{\dim \mathcal H _L - 1},
\end{align}
and where $\Lambda_L$ is the logical channel which acts on $\rho_L$ as 
\begin{align} \label{eq:logical_channel} 
\Lambda_L[\rho_L] := \Tr_{\rm synd}\big [\mathcal{U}^{-1} \mathcal D  \Recb \widetilde{\mathcal{U}} \mathcal E (\rho_L \otimes \op{0}{0}_{\rm synd}) \big].
\end{align}
Importantly, the dimension dependence of the decay parameter only involves the dimension of the logical space. 

Our protocol strictly generalizes the protocol of \autoref{eq:LogFid}
by including the effect of state preparation and measurement (SPAM)
errors. We note that $F_L(m=1) = F_L( \mathcal{R} \Lambda)$ 
in the absence of SPAM,
such that the logical sequence and gate fidelities coincide for sequences of
length 1. In the case of non-trivial SPAM errors, one can still estimate gate fidelity 
by estimating the
logical sequence fidelity for sequences of several different lengths.
Since each length of sequence depends on the SPAM errors differently,
this allows us to isolate the contribution of the SPAM errors.
For example,
if $T = 1$, then $q_i$ is a Bernoulli random variable such that $\overline{q}(m)$ is binomially-distributed, and admits an efficient algorithm for estimating $F_L(\tilde{\mathcal{R}}\overline{\Lambda})$
\cite{GranFerrCory15}.

In the case of performing our protocol without a recovery operation
($\mathcal{R} = \id$), we obtain an estimate of $F_L(\overline{\Lambda})$.
Recalling \autoref{eq:no_err}--\autoref{eq:un_err}, we can then estimate averaged code
properties by comparing these estimates.
For instance, we can estimate
\begin{subequations}
    \begin{align}
        \widehat{\Pr}(\Un | \mathcal{R}, \overline{\Lambda}) =
        \expect_\psi\left[\widehat{\Pr}(\Un | \psi, \mathcal{R}, \overline{\Lambda}) \right] & =
            1 - \hat{F}_L(\tilde{\mathcal{R}}\overline{\Lambda}), \\
        \widehat{\Pr}(\Co | \mathcal{R}, \overline{\Lambda}) =
        \expect_\psi\left[\widehat{\Pr}(\Co | \psi, \mathcal{R}, \overline{\Lambda}) \right] & =
            \hat{F}_L(\tilde{\mathcal{R}}\overline{\Lambda}) - \hat{F}_L(\overline{\Lambda}),\\
              \text{and }             \widehat{\Pr}(\No | \mathcal{R}=\id, \overline{\Lambda}) =
    \expect_\psi\left[\widehat{\Pr}(\Co | \psi, \id, \overline{\Lambda}) \right] & =
     \hat{F}_L(\overline{\Lambda}),
    \end{align}
\end{subequations}
where we have used $\widehat{\Pr}$ to denote that each of these quantities
is our estimate of the relevant code properties given data collected from our
logical RB protocol.

We thus see that our protocol allows us to efficiently and robustly
estimate several different averaged properties of
the error correcting code implementation under study,
and to compare these properties for different choices
of recovery $\mathcal{R}$. 
Undoubtedly, the protocol we suggest is demanding for current experiments, but many simplifications can be made if we consider a stabilizer code with Pauli preparations and measurements, which is by far the most commonly considered case. 
Under these assumptions, we need not actually perform the recovery operation, as it is always a Pauli operation that we can commute forward through the rest of the circuit. 
Thus, we can also assume a perfect recovery in the case of stabilizer codes. 
This is understood as follows. 
The recovery ${R}(s)$ for a syndrome bitstring $s$ can then be propagated
through each subsequent logical Clifford operation and the final encoder $E$ using Gottesman--Knill simulation
to obtain an equivalent recovery operator ${R'}(s)$ that can be applied immediately
before measurement.
Each such recovery then either flips or does not flip the result of measuring survival,
by the structure of the Pauli group. If an even number of propagated recovery operators flip the resulting
measurement, then a survival is recorded as normal, while if an odd number flip, the roles of
survival and anti-survival are exchanged for that measurement.
Critically, this reasoning is a function of the syndromes $s_j$ collected after the $j$th encoded
gate and of the sequence $\vec{i}$ performed, such that \emph{no particular recovery map needs to
be chosen at experiment time}. Choosing the trivial recovery ${R}(s) = {R'}(s) = {\id}$, for instance,
will report the logical fidelity without the aid of recovery operators. By comparing
a particular nontrivial recovery map to these results, LRB isolates the performance not only
of the encoded gate, but also of the ensuing recovery operator and syndrome measurement. 

\section{Advantages of Logical Characterization} 

There are multiple reasons why one should favor LRB, and in general logical characterization methods, over physical RB. We now elucidate on two such reasons. First, physical characterization is a bad predictor of logical performance, sometimes dramatically so~\cite{Poulin2017Coogee}. 
Here we construct examples that demonstrate how physical RB will in general under or over estimate logical fidelity.
Second, logical RB is significantly more efficient than physical characterization on
the full register of interest, as we discuss in detail below.

\subsection{LRB honestly assesses code performance}
Let us begin with the case where physical RB overestimates the logical fidelity. We consider the example of a $\llbracket3, 1, 1\rrbracket$ bit-flip
 code with stabilizers $S = \langle ZZ\id, \id ZZ\rangle$, with logical operators $\overline{X} = XXX$ and $\overline{Z} = ZZZ$, and with perfect encoding and decoding, i.e.\ $\mathcal \En = \id \mathcal E $ and $\mathcal \De = \id \mathcal D$.
As we have shown that logical RB provides a robust estimate of average gate fidelities,
in this section we will focus on what we learn from gate fidelity and will hence temporarily
ignore SPAM.
 
 \begin{figure}[t]
    \begin{center}
        \includegraphics[width=0.6\columnwidth]{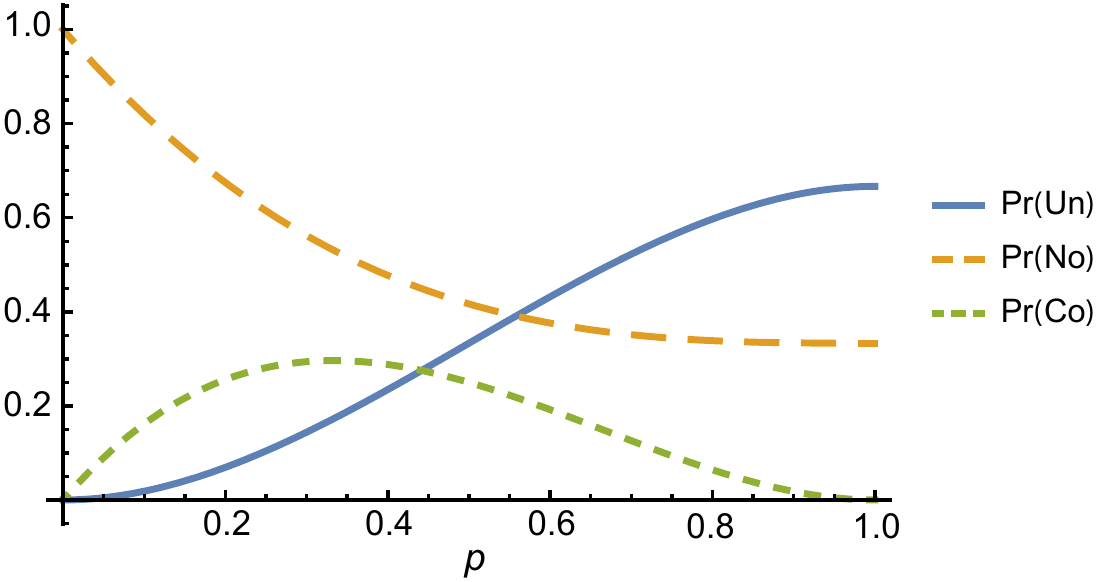}
        \caption{
            \label{fig:pr_err}
            The error probabilities for a separable channel given in \autoref{eq:prob_example}. The maximum of $\Pr(\Co)$ occurs at $p=1/3$, as expected, while $\Pr(\Co)=\Pr(\Un)$ when $p= \left(9-\sqrt{21}\right)/10\approx 0.441$.
        }
    \end{center}
\end{figure}
 
The physical error map that we use is the composition of independent bit-flip 
errors and pairwise correlated errors between qubits $\overline{\Lambda}= \overline{\Lambda}^{\rm cor} \overline{\Lambda}^{\rm ind}$. The independent channel acts as $ \overline{\Lambda}^{\rm ind}(\rho) = \mathcal E_3 \mathcal E_2 \mathcal E_1(\rho) $, where $\mathcal E_i (\rho) = (1-p) \rho + p X _i \rho X_i$ and we denote composition of channels by multiplication. 
The correlated channel acts as $ \overline{\Lambda}^{\rm cor}(\rho) =  \mathcal E_{2,3} \mathcal E_{1,2}$ where $\mathcal E_{i,j}(\rho)= (1-q)\rho + q X_i X_j  \rho X_i X_j $ and the subscripts $i,j$ denote which qubit(s) the channel acts on; for all other qubits there is an implied identity. In the absence of correlated errors
 ($q = 0$), the logical fidelity with and without recovery is given by \citet{RahnDoheMabu02} as
\begin{subequations}
 \begin{align}
     F_{L,\text{rec}}(p) & \defeq F_{L}^{\rm ind}(\mathcal{R} \overline{\Lambda}) = 1 - 2 p^2 + \frac43 p^3\\
     F_{L,\text{no rec}}(p) & \defeq F_{L}^{\rm ind }(\overline{\Lambda}) = 1 - 2p +2p^2 - \frac23 p^3.
      \end{align}
\end{subequations}
These translate into the average error probabilities
 \begin{subequations}\label{eq:prob_example}
 \begin{align}
     \Pr(\No) &= F_{L,\text{no rec}}^{\rm ind }(p)\\
     \Pr(\Co) &=  F_{L,\text{rec}}^{\rm ind}(p) -F_{L,\text{no rec}}^{\rm ind }(p) \\
     \Pr(\Un) &=  1-F_{L,\text{rec}}^{\rm ind }(p),
      \end{align}
\end{subequations}
as we illustrate in \autoref{fig:pr_err}.
The logical fidelity for the total channel ($q \ne 0$) is
 \begin{subequations}
\begin{align}
F_{L,\text{rec}}(p, q) &= F_{L,\text{rec}}^{\rm ind}(p) -\frac{4 }{3}q+\frac{8 }{3}p q+\frac{2 }{3}q^2 -\frac{1}{3} 4 p q^2\\
F_{L,\text{no rec}}(p, q) &=  F_{L,\text{no rec}}^{\rm ind }(p)-\frac{4 }{3}q+4 p q +\frac{2 }{3}q^2-\frac{8 }{3}p^2 q -2 p q^2
+\frac{4 }{3}p^2 q^2,
\end{align}
\end{subequations}
and the probabilities have the same form as \autoref{eq:prob_example}.

We first imagine performing physical RB on one of the three qubits, which involves tracing over the other qubits. The estimated fidelity is
 $F_{P}^{\rm est}(p,q)=1-\frac{2 }{3}p-\frac{2 }{3}q+\frac{4 }{3}p q$, while for a truly independent channel it is $F_{P}^{\rm ind}(p)=1-\frac{2 }{3}p$. 
Using the standard assumption of independent noise channels, one would infer a physical error rate given in terms of this estimated fidelity of $p^{\rm est} = \frac{3}{2}[1-F_{P}^{\rm est}(p,q)]$. 
 
\begin{figure}[t!]
    \begin{center}
        \includegraphics[width=0.7\columnwidth]{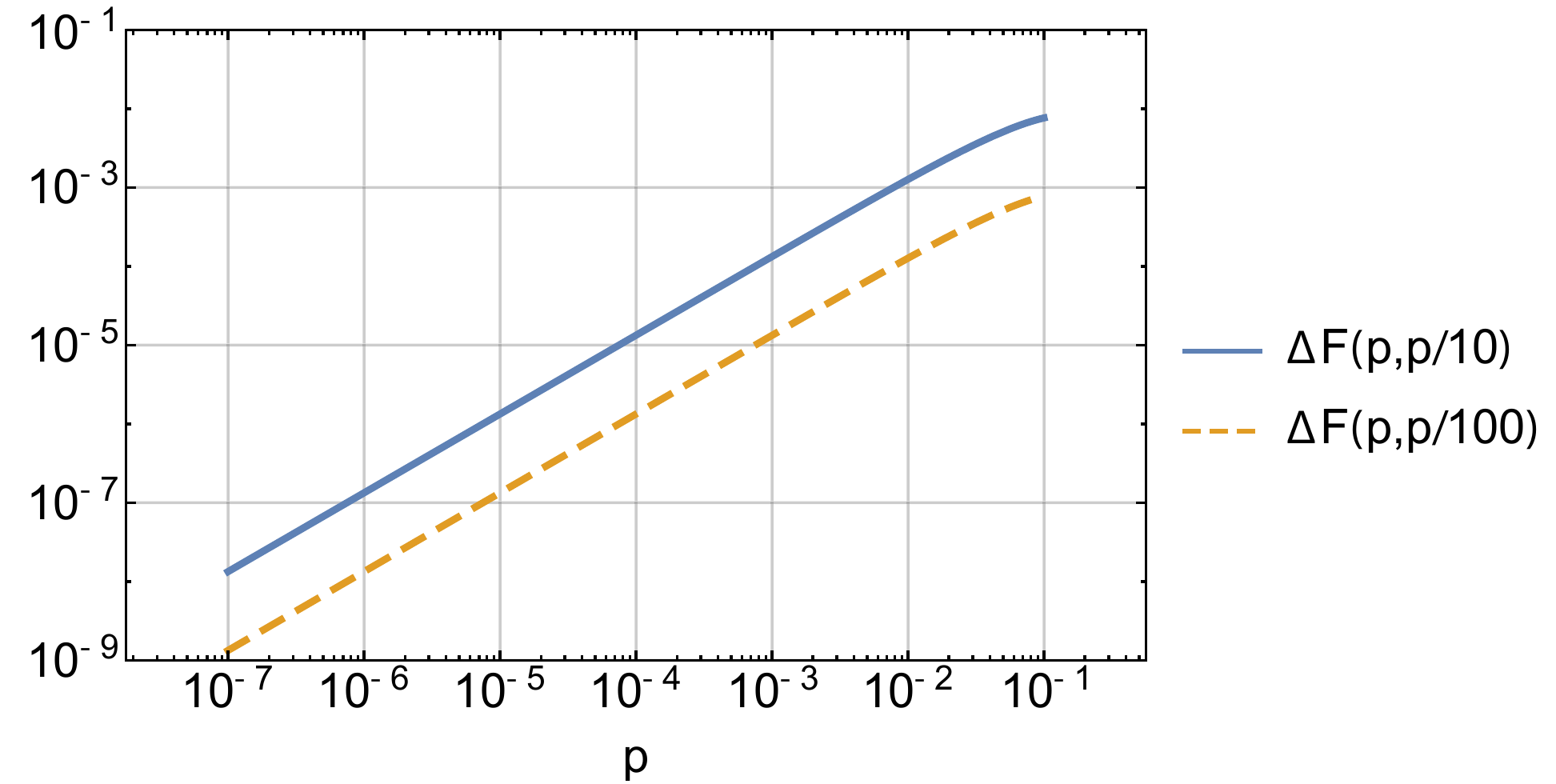}
        \caption{
            \label{fig:delta_f}
            Overestimation of logical fidelity when the physical noise channel
            is correlated and the recovery operation is performed. We plot the
            difference between the true fidelity and the estimated fidelity
            $\Delta F_L(p,q)$, where $p$ is the uncorrelated noise parameter
            and $q$ is the correlated noise parameter. We consider two cases:
            $q = p / 10$ or $q = p /100$. That is, the correlated noise
            parameter $q$ is taken to be one or two orders of magnitude
            smaller than $p$. To leading order in $p$ and $q$, we find that
            $\Delta F_L(p, q) \approx (4/3)q$, which is a good approximation
            for $p, q< 10^{-2}$. This asymptotic analysis and the plot of $\Delta
            F_L(p,q)$ indicate the estimated logical fidelity will be
            incorrect at same order of magnitude as the correlation parameter
            $q$.
        }
    \end{center}
\end{figure}

 Estimating the logical fidelity using $p^{\rm est}$ assuming no correlations, i.e.\ $F_L^{\rm ind}(p^{\rm est})$, will then return an \emph{overestimate} of the actual logical fidelity. This can be illustrated by considering the difference between the estimated and true fidelity $\Delta F_L(p,q) = F_L^{\rm ind}(p^{\rm est}) -F_L(p, q)$ which is positive (an over estimate) for $p,q\in [0,0.5]$, see \autoref{fig:delta_f}.
 As a consequence of the fact that the code properties are linearly related to $\Delta F_L(p, q)$,
 the estimated probabilities obtained from physical RB will also be incorrectly inferred.

The case when physical RB underestimates the logical fidelity is much simpler. 
First, suppose only one qubit experiences a bit flip channel while all others are perfect, and then physical RB is performed on that qubit. 
Clearly $F_L^{\rm ind}=1$, as the bit flip code can correct this error perfectly, but the physical fidelity $F_{P}^{\rm ind}(p)=1-\frac{2 }{3}p$ is less than 1. 
This simple example extends to a more natural and symmetric error channel that has anti-correlated noise. 
Consider an equal convex combination of independent bit flip channels, i.e.\ $\mathcal{F}(\rho)= \tfrac{1}{3} \bigl[\mathcal{E}_3(\rho) + \mathcal{E}_2(\rho) + \mathcal{E}_1(\rho)\bigr]$. 
Again $F_L^{\rm ind}=1$, as the bit flip code can correct any single bit-flip error. 
However, the physical fidelity $F_{P}^{\rm ind}(p)=1- \frac{2}{9} p$ will be extrapolated to $F_L^{\rm ind}(p)= 1 -2 (p/3)^2 +(4/3) (p/3)^3$ for the logical fidelity---again less than one. 

While we have focused on LRB, the problems we illustrate are generic for physical vs.\ logical characterization.
There are many ways such channels could arise, including correlated errors, control errors, crosstalk and more.  
Any one of these examples gives a strong reason for direct logical characterization.  

\subsection{LRB is more efficient than physical characterization}

In principle,
one could use an entire diagnostic suite at the physical level to deal with each problem raised above,
including estimating the unitarity~\cite{WallGranHarp15}, crosstalk~\cite{GambCorcMerk12}, and other extensions to RB~\cite{MageGambJohn12,KimmSilvRyan14}.
This suffers a very large overhead, however, due to the fact that one must know quite a lot
about the physical performance of a system in order to predict its logical performance.
For instance, current approaches to mitigating the problems described in the previous section
rely on using very large models such as gate set tomography to make rigorous claims about
error rates and fault tolerance \cite{BlumGambNiel16}.

This complexity is also manifest in physical RB directly,
as one needs to both sample from the $n$-qubit Clifford group, and then
decompose each sampled Clifford gate in terms of an elementary gate set. 
For the most commonly used Clifford gate set, this costs $O(n^2/\log n)$ elementary gates. By contrast, in LRB,
we only require the $k$-qubit Clifford group compiled at the logical level. As noted above, the recovery
can be done in post-processing for stabilizer codes, such that the quadratic overhead does not impose
any significant experimental limitations beyond that already incurred by implementing error correction
at all.

By concentrating instead on the logical degrees of freedom, LRB effectively uses the structure
of an error correcting code to remove degrees of freedom which do not directly affect the performance
of that code. Moreover, because our method reduces to conventional RB, logical RB immediately
also implies that any relevant subset of the full diagnostic suite built up from conventional RB can also be used
at the logical level, without mandating that the entire suite be used. For instance, the purity benchmarking
protocol can be used at the logical level to test hypotheses about how the coherences of errors
transform under error correction by reporting logical unitarities \cite{WallGranHarp15,FengBuonWall16}. Similarly, LRB can be used
together with crosstalk benchmarking~\cite{GambCorcMerk12} to test if any correlations persist between logical
qubits, or with leakage benchmarking~\cite{WallBarnEmer14} to test leakage of the encoded information to outside the code space. 
Finally, using dihedral \cite{CrosMageBish16,DugaWallEmer15}
or CTPT \cite{HarpFlam16} RB
at the logical level, we can assess the quality of $T$ gates injected into a
candidate for fault-tolerant quantum information processing.

\section{Conclusion} 

We have described a protocol to measure averaged versions of the three most important properties of an error correction
implementation: the average probability of correctable errors and uncorrectable errors as well as the average error rate of a gate set. 
The strengths of LRB are three-fold: Firstly, it does not under- or overestimate errors at the logical level, which is a significant advantage since physical characterization and extrapolation can lead to misleading conclusions.  
Furthermore, the protocol itself fits naturally within the framework of standard RB theory by reducing to the most well-studied case of standard RB on the logical degree of freedom. 
This enables us to carry over the known results at the physical level from the full suite of RB diagnostic tools directly into the logical level. 
Finally, it achieves a drastic reduction in the experimental resources required to characterize an error-corrected quantum device. 
The latter strength is for two reasons. First, it reduces the Hilbert space dimension by acting at the logical level. 
Secondly---in the case of stabilizer codes---the correction operation can be applied in post-processing rather than concurrently.

In this work we have not focused on studying the totality of the information contained in the syndrome measurements. 
This has been explored by \citet{CombFerrCesa14} where it was shown that some additional information about the error channel is contained in the statistics of the syndrome measurements. 
It is an open question if the statistics of syndrome measurements used in our LRB characterization method can provide additional information about the logical or physical error channels.

It would also be interesting to investigate time-correlated errors such as have been studied by \citet{EpstCrosMage14} and \citet{BallStacFlam16}. 
When this type of noise is generated by slowly varying control fields (low-frequency noise), this can lead to qualitatively different behavior in RB experiments. 
However, there is evidence that this type of noise becomes less coherent at the logical level~\cite{GutiSmitLulu16}, so it is not unreasonable to hope that LRB actually performs better and more reliably than its physical counterpart in this respect. 

We expect that our protocol can be performed in the near future. Indeed, the
recent experimental work of \citet{HeerReinOfek16} implemented a protocol
which contains many elements of our LRB protocol. There are only a few
elements which need to be added to implement our LRB proposal.  In their
experiment, a transmon qubit is prepared in some state and then that state is
encoded it into an oscillator (a cavity) in a cat-state code~\cite{mirrahimi_dynamically_2014}. 
Then they perform sequences of Clifford gates on the logical qubit, reporting an logical
RB decay of $\hat{p}_L = \exp(-1/\hat{\tau})\approx 0.982$ where
$\hat{\tau}=55\pm 0.7$. Their protocol differs from ours, however, in that the
Heeres \etal~protocol does not use recovery operations or syndrome resets,
thus we cannot directly conclude $\Pr(\No)\approx 0.982$.

We close by mentioning a possible application of our protocol. One could imagine experimentally searching for an optimal code---minimizing the probability of an uncorrectable error---over a class of codes for a given channel without actually knowing what the channel is. 
To be specific, consider an error channel which is some combination of unitary and stochastic errors on five qubits encoded in the five qubit code. 
Then we may use LRB to determine the probability for an uncorrectable error. Next, we can unitarily rotate the code space of the five qubit perfect code and determine the probability for an uncorrectable error again, as was recently suggested by \citet{FlorBrun16}. 
By iterating this procedure and searching over unitary rotations of the code space, LRB can allow a stochastic optimization procedure to minimize the probability for an uncorrectable error for the unknown channel.


\acknowledgments{
    We thank Christopher Chamberland, Tomas Jochym-O'Connor, David Poulin, and Joel Wallman for discussions.
    JC acknowledges support in part by the Perimeter Institute for Theoretical Physics. 
    Research at Perimeter Institute is supported
    by the Government of Canada through the Department
    of Innovation, Science and Economic Development
    Canada and by the Province of Ontario through
    the Ministry of Research, Innovation and Science.
    This work was also supported by the Australian Research Council via EQuS project number CE11001013, and 
    by the US Army Research Office grant numbers W911NF-14-1-0098 and W911NF-14-1-0103. 
    JC and STF were supported by the Australian Research Council via DE160100356 and via FT130101744 
    respectively.
}

\nocite{apsrev41Control}
\bibliographystyle{apsrev4-1}
\bibliography{apsrev-control,logical_rb_bib}


\appendix
\onecolumngrid

\section{Appendix: The Derivation of Logical Randomized Benchmarking}\label{apx:derivation}
In this appendix we begin by revisiting, in section \autoref{apx:prb}, the derivation of Physical RB using 
the graphical derivation of \citet{GranFerrCory15}.  Then in section \autoref{apx:ancilla_coupled}
we show how a logical channel derived from ancilla coupled measurements can be reduced to an effective
direct measurement channel. Finally, in section \autoref{apx:LRB_derivation}, we derive the logical randomized benchmarking protocol by reduction to
the 0th-order model of \citet{MageGambEmer12} using the graphical techniques of \citet{GranFerrCory15}.

\subsection{Physical RB}
\label{apx:prb}

\newcommand{\LP}{\ensuremath{\Lambda_{\mathrm{P}}}}
\newcommand{\LM}{\ensuremath{\Lambda_{\mathrm{M}}}}

Consider circuits of the form
\begin{subequations}
    \begin{align}
        Q\Lambda_{\rm M} \Lambda U^{-1} \Lambda U_m \cdots \Lambda U_2 \Lambda U_1 \ket{\psi} = & 
        \raisebox{-0.65em}{
            \tikz[baseline=-0.5ex]{
                \node[] at (-9.125, 0.25) {$\ket{\psi}$};
                \draw[pwire] (-12.5, 0.25) -- (-9.5, 0.25);
                \draw[pwire] (-15.5, 0.25) -- (-13.5, 0.25);
                \node[gate, physical, minimum height=1.5em] at (-10, 0.25) {$\LP$};
                \node[gate, physical, minimum height=1.5em] at (-11, 0.25) {$\widetilde U_1$};
                \node[gate, physical, minimum height=1.5em] at (-12, 0.25) {$\widetilde U_2$};
                \node at (-13, 0.25) {$...$};
                \node[gate, physical, minimum height=1.5em] at (-14, 0.25) {$\widetilde U^{-1}$};
                \node[gate, physical, minimum height=1.5em] at (-15, 0.25) {$\Lambda_{\rm M}$};
                \node[] at (-15.75, 0.25) {$Q$};
            }
        } \\ 
        = & 
        \raisebox{-0.65em}{
            \tikz[baseline=-0.5ex]{
                \node[] at (-6.125, 0.25) {$\ket{\psi}$};
                \draw[pwire] (-11.5, 0.25) -- (-6.5, 0.25);
                \draw[pwire] (-15.5, 0.25) -- (-12.5, 0.25);
                \node[gate, physical, minimum height=1.5em] at (-7, 0.25) {$\LP$};
                \node[gate, physical, minimum height=1.5em] at (-8, 0.25) {$U_1$};
                \node[gate, physical, minimum height=1.5em] at (-9, 0.25) {$\Lambda$};
                \node[gate, physical, minimum height=1.5em] at (-10, 0.25) {$U_2$};
                \node[gate, physical, minimum height=1.5em] at (-11, 0.25) {$\Lambda$};
                \node at (-12, 0.25) {$...$};
                \node[gate, physical, minimum height=1.5em] at (-13, 0.25) {$U^{-1}$};
                \node[gate, physical, minimum height=1.5em] at (-14, 0.25) {$\Lambda$};
                \node[gate, physical, minimum height=1.5em] at (-15, 0.25) {$\LM$};
                \node[] at (-15.75, 0.25) {$Q$};
            }
        }\!,
        \label{eq:reduction-target-eq}
    \end{align}
\end{subequations}
where $Q$ is a measurement effect (typically it is taken to be a measurement of $\op{\psi}{\psi}$ vs $\id-\op{\psi}{\psi}$), $\ket{\psi}$ is a preparation,
$\LM$ and $\LP$ are channels describing errors in the measurement
and preparation respectively,
$\{U_i\}$ are a sequence of $m$ randomly chosen Clifford gates with inverse
$U^{-1} \defeq \left(\prod_{i = m}^1 U_i \right)^{-1}$,
and where $\Lambda U_i$    
is the imperfect implementation of the unitary gate $U_i$.
Pictorially, we represent this by a circuit diagram using the Kitaev convention of time $t$ increasing to the left.
 
The result of \citet{MageGambEmer12} is then that
the expectation value over sequences results in a simple likelihood function
\begin{align}
    \label{eq:magesan-likelihood}
    \Pr(\text{survival} | A, B, p; m) = \expect_{U_1, \dots, U_m}[ \Tr (
     Q \Lambda U^{-1} \Lambda U_m \cdots \Lambda U_2 \Lambda U_1 \rho)
    ] & =
    A p^m + B,
\end{align}
where
\begin{subequations}
 \label{eq:zeroth-order-defns}
 \begin{align}
     A & \defeq \Tr(E_\psi \Lambda[\rho_\psi - \id / d]), \\
     B & \defeq \Tr(E_\psi \Lambda[\id / d]), \\
     p & \defeq (d F(\Lambda) - 1) / (d - 1), \\
     Q_\eff & \defeq \Lambda_M\dg(Q)\text{, and} \label{subeq:effective-phys-rb-meas} \\
     \rho_\psi &\defeq  \Lambda_P(\op{\psi}{\psi}).
 \end{align}
\end{subequations}
Following~\citet{GranFerrCory15}, we can graphically summarize the
proof of \citet{MageGambEmer12} by noting that we can use a change of
variables $V_1 = U_1$ and $V_i \defeq U_{i} V_{i - 1}$ 
to rewrite the expectation over sequences. Concretely, making this
change of variables, we obtain that
\begin{gather}
    \begin{aligned}
        \label{eq:prb-deriv}
        & \expect_{\vec{i}}\Big[
            \tikz[baseline=-0.5ex]{
                \node at (-10, 0) {$Q_\eff$};
                \draw[pwire] (-8, 0) -- (-9.5, 0);
                \node[gate, physical, minimum height=1.5em] at (-9, 0) {$\Lambda$};
                \draw[pwire] (-7, 0) -- (-8, 0);
                \node[gate, physical, minimum height=1.5em] at (-8, 0) {$U^{-1}$};   
                \draw[pwire] (-5, 0) -- (-7, 0);
                \node[gate, physical, minimum height=1.5em] at (-6, 0) {$\Lambda$};
                \draw[pwire] (-4, 0) -- (-5, 0);
                \node[gate, physical, minimum height=1.5em] at (-5, 0) {$U_{m}$};                 
                \node[] at (-3, 0) {$\cdots$};
                \draw[pwire] (-2, 0) -- (0, 0);
                \node[gate, physical, minimum height=1.5em] at (-1, 0) {$\Lambda$};
                \draw[pwire] (0, 0) -- (2, 0);
                \node[gate, physical, minimum height=1.5em] at (0, 0) {$U_2$};
                \draw[pwire] (2, 0) -- (3, 0);
                \node[gate, physical, minimum height=1.5em] at (2, 0) {$\Lambda$};
                \draw[pwire] (3, 0) -- (3.5, 0);
                \node[gate, physical, minimum height=1.5em] at (3, 0) {$U_1$};
                \node at (3.875, 0) {$\rho_\psi$};
            }        
        \Big] \\
        = {} & \expect_{\vec{i}}\Big[
            \tikz[baseline=-0.5ex]{
                \node at (-10, 0) {$Q_\eff$};
                \draw[pwire] (-8, 0) -- (-9.5, 0);
                \node[gate, physical, minimum height=1.5em] at (-9, 0) {$\Lambda$};
                \draw[pwire] (-7, 0) -- (-8, 0);
                \draw[pwire] (-5, 0) -- (-7, 0);
                \node[gate, physical, minimum height=1.5em] at (-7, 0) {$V_{m}\dg$};   
                \node[gate, physical, minimum height=1.5em] at (-6, 0) {$\Lambda$};
                \draw[pwire] (-4, 0) -- (-5, 0);
                \node[gate, physical, minimum height=1.5em] at (-5, 0) {$V_{m}$};                 
                \node[] at (-3, 0) {$\cdots$};
                \draw[pwire] (-2, 0) -- (0, 0);
                \node[gate, physical, minimum height=1.5em] at (-2, 0) {$V_2\dg$};
                \node[gate, physical, minimum height=1.5em] at (-1, 0) {$\Lambda$};
                \draw[pwire] (0, 0) -- (2, 0);
                \node[gate, physical, minimum height=1.5em] at (0, 0) {$V_2$};
                \draw[pwire] (2, 0) -- (3, 0);
                \node[gate, physical, minimum height=1.5em] at (1, 0) {$V_1\dg$};
                \node[gate, physical, minimum height=1.5em] at (2, 0) {$\Lambda$};
                \draw[pwire] (3, 0) -- (3.5, 0);
                \node[gate, physical, minimum height=1.5em] at (3, 0) {$V_1$};
                \node at (3.875, 0) {$\rho_\psi$};
            }    
        \Big] \\
        = {} & \hphantom{\expect_{\vec{i}}\Big[}
        \tikz[baseline=-0.5ex]{
            \node at (-10, 0) {$Q_\eff$};
            \draw[pwire] (-8, 0) -- (-9.5, 0);
            \node[gate, physical, minimum height=1.5em] at (-9, 0) {$\Lambda$};
            \draw[pwire] (-7, 0) -- (-8, 0);
            \draw[pwire] (-5, 0) -- (-7, 0);
            \node[gate, physical, minimum height=1.5em,minimum width=4em] at (-6, 0) {\!\!$W[\Lambda]$};
            \draw[pwire] (-4, 0) -- (-5, 0);
            \node[] at (-3, 0) {$\cdots$};
            \draw[pwire] (-2, 0) -- (0, 0);
            \node[gate, physical, minimum height=1.5em,minimum width=4em] at (-1, 0) {\!\!$W[\Lambda]$};
            \draw[pwire] (0, 0) -- (2, 0);
            \draw[pwire] (2, 0) -- (3, 0);
            \node[gate, physical, minimum height=1.5em,minimum width=4em] at (2, 0) {\!\!$W[\Lambda]$};
            \draw[pwire] (3, 0) -- (3.5, 0);
            \node at (3.875, 0) {$\rho_\psi$};.
        },
    \end{aligned}
\end{gather}
where $W[\Lambda]$ is the twirling superchannel acting on $\Lambda$,
defined as
\begin{align*}
    W[\Lambda](\rho) \defeq p \rho + (1 - p) \frac{\id}{d}.
\end{align*}
Note that in the second line, we have applied the change of variables $U \to V$ in order to cancel out
the previous $U_{i-1}$ gates.
In the third line, we have used that because we chose each $U_i$ randomly from the Clifford group,
the $V$ gates then also form a two-design.
On the last line the expectation value over random gates in the previous two lines has been taken. This gives a twirling superchannel $W$ acting on the noise $\Lambda$.

\subsection{Ancilla-coupled Logical Fidelity Measurement}
\label{apx:ancilla_coupled}

In the main text, we considered only \emph{in-place} error correction, and did not consider ancilla-coupled syndrome measurements.
Though this allows for a more simple presentation of our protocol, we note that ancilla-coupled
syndrome measurements are likely to be used in practical applications of quantum error correction.
Thus, we will now detail how to reduce ancilla-coupled protocols to those described
in the main text.

To do so, we start with a more complicated circuit and reduce it to circuit \autoref{cir:logfid} in the main text.  
The state $\ket{\psi}$  is encoded into a quantum code
using a noisy encoding operation $\mathcal\En=\mathcal E \Lambda_P$ and eventually noisly decoded
$\mathcal \De= \Lambda_M \mathcal D $, these operations represent SPAM errors.
A noisy logical unitary
$\widetilde{\mathcal U}$ is applied to the encoded state, then a measurement of the syndromes is 
achieved by a noisy coupling $\widetilde A$ to some ancilla. Then a measurement 
of projectors on to the syndromes $\Pi_s$ 
is made nonideal by including channels for 
state preparation $\Lambda_P$ and measurement $\Lambda_M$ errors.
These syndromes are used to enact a noisy recovery operation $\Recb'(s)$.
 After noisily decoding and perfectly inverting $U$ we measure $\Meaz$ vs $\Meazb$. This procedure 
can be implemented by the circuit 

\begin{center}
    \tikz[baseline=-1\baselineskip]{ 
        \begin{scope}[every node/.style={scale=0.86}, scale=0.86]
            \node[] at (1, 1.0) {$\ket{\psi}$};
            \node[] at (1, 0.5) {$\ket{0}$};
            \node[] at (1, 0.0) {$\ket{0}$};
            \node[] at (1, -0.5) {$\ket{0}$};
            \node[] at (1, -1.0) {$\ket{0}$};
            \draw[pwire] (0.5, 1.0) -- (-8, 1.0);
            \draw[pwire] (0.5, 0.5) -- (-8, 0.5);
            \draw[pwire] (0.5,  0.0) -- (-8, 0.0);
            \draw[pwire] (0.5, -0.5) -- (-4, -0.5);
            \draw[pwire] (0.5, -1.0) -- (-4, -1.0);
            \node[gate, physical, minimum height=4em] at (0, 0.5) {$\mathcal\En$};
            \node[gate, physical, minimum height=2.6em] at (0, -0.75) {$\Lambda_P$};
            \node[gate, physical, minimum height=4em] at (-1, 0.5) {$\widetilde{\mathcal U}$};
            \node[gate, physical, minimum height=6.6em] at (-2, 0.0) {$\widetilde A$};
            \node[gate, physical, minimum height=2.6em] at (-3, -0.75) {$\Lambda_M$};
            \draw[pwire, cwire] (-4.4, -0.75) -- (-5, -0.75) -- (-5, 0.25);
            \node[meas, physical, scale=0.8] at (-4, -0.75) {$s$};
            \node[gate, physical, minimum height=4em,minimum width=3em] at (-5, 0.5) {\!$\Recb'(s)$};
            \node[gate, physical, minimum height=4em] at (-6, 0.5) {$\mathcal\De$};
            \node[gate, physical, minimum height=1.3em,minimum width=2em] at (-7, 1) {$U^{-1}$};
            \node[meas, physical, scale=0.7] at (-8, 1) {$\psi$};
            \node[meas, physical, scale=0.8] at (-8, 0.25) {${\rm Tr}$};
        \end{scope}
    }.
\end{center}

We will \emph{assume} that we can trace out the syndrome qubits, which means they can not be correlated between rounds of error correction. This assumption is identical to the assumption made in the main text that
we can refresh ancilla qubits such that any errors are uncorrelated with their previous
states; that is, we need not introduce additional assumptions to reason about
the case of ancilla-coupled measurements.
Thus, we have an equivalent representation without ancilla, given
by the quantum operation 
$ \Recb \odot = \Tr _{\rm ancilla}\big [\Recb ' \mathcal P_{s} \Lambda_M \widetilde A \Lambda_P \op{0}{0}_{\rm ancilla}\odot \big ]$,
where $\odot$ is a placeholder and $\mathcal P_s$ is a quantum operation representation of $\Pi_s$. 
Finally, the above circuit becomes
\begin{center}
    \tikz[baseline=0\baselineskip]{ 
        \begin{scope}[every node/.style={scale=0.86}, scale=0.86]
            \node[] at (1, 1.0) {$\ket{\psi}$};
            \node[] at (1, 0.5) {$\ket{0}$};
            \node[] at (1, 0.0) {$\ket{0}$};
            \draw[pwire] (0.5, 1.0) -- (-5, 1.0);
            \draw[pwire] (0.5, 0.5) -- (-5, 0.5);
            \draw[pwire] (0.5,  0.0) -- (-5, 0.0);
            \node[gate, physical, minimum height=4em] at (0, 0.5) {$\mathcal\En$};
            \node[gate, physical, minimum height=4em] at (-1, 0.5) {$\widetilde U$};
            \node[gate, physical, minimum height=4em] at (-2, 0.5) {$\Recb$};
            \node[gate, physical, minimum height=4em] at (-3, 0.5) {$\mathcal\De$};
            \node[gate, physical, minimum height=1.3em,minimum width=2em] at (-4, 1) {$U^{-1}$};
            \node[meas, physical, scale=0.7] at (-5, 1) {$\Meaz$};
            \node[meas, physical, scale=0.8] at (-5, 0.25) {${\rm Tr}$};
        \end{scope}
    }.
\end{center}
We recognize this as being identical to \autoref{cir:logfid}, completing the
reduction.

\subsection{Logical RB}
\label{apx:LRB_derivation}

We would now like to derive expressions of the form \autoref{eq:magesan-likelihood} and circuits of the form
\autoref{eq:prb-deriv} for our logical randomized benchmarking protocol as described in the main body.
To do so, we start with the physical circuit
\begin{equation} 
    \tikz[baseline=-1em, every node/.style={font=\relsize{-1}}] {
        \draw[pwire] (-1.5, 0.25) -- (4, 0.25);
        \draw[pwire] (-1.5, -0.25) -- (4, -0.25);
        \draw[pwire] (-1.5, -0.75) -- (4, -0.75);
        \draw[pwire] (-8, 0.25) -- (-2.5, 0.25);
        \draw[pwire] (-8, -0.25) -- (-2.5, -0.25);
        \draw[pwire] (-8, -0.75) -- (-2.5, -0.75);
        \node[gate, physical, minimum height=4em] at (3, -0.25) {$\En$};
        \node[gate, physical, minimum height=4em] at (2, -0.25) {$\widetilde U_1$};
        \node[gate, physical, minimum height=4em] at (1, -0.25) {$\Recb$};
        \node[gate, physical, minimum height=4em] at (0, -0.25) {$\widetilde U_2$};
        \node[gate, physical, minimum height=4em] at (-1, -0.25) {$\Recb$};
        \node[] at (-2, -0.25) {$\cdots$};
        \node[gate, physical, minimum height=4em] at (-3, -0.25) {$\widetilde U_m$};
        \node[gate, physical, minimum height=4em] at (-4, -0.25) {$\Recb$};
        \node[gate, physical, minimum height=4em] at (-5, -0.25) {$\widetilde{ U}^{-1}$};
        \node[gate, physical, minimum height=4em] at (-6, -0.25) {$\Recb$};
        \node[gate, physical, minimum height=4em] at (-7, -0.25) {$\De$};
        \node[] at (4.5, 0.25) {$\ket{\psi}$};
        \node[] at (4.5, -0.25) {$\ket{0}$};
        \node[] at (4.5, -0.75) {$\ket{0}$};
        \node[] at (-8.75, 0.25) {$Q$ vs $\bar Q$};
        \node[] at (-8.75, -0.5) {Tr};
    }.
\end{equation}
It is sufficient to consider $U_1,U_2,U^{-1}$, so we specialize to that case
by deriving our reduction for sequences of length $m=2$.
Next, we expand the imperfect gates (denoted by quantum operations with overset tildes)
by decomposing them into perfect operations followed by noise channels.
For example, suppose $\overline{\mathcal{X}}$ is an ideal quantum operation; then,
we denote its imperfect implementation as
$\widetilde{\mathcal{X}}=\overline{\Lambda}_{\mathcal X} \overline{\mathcal{X}}$.
With this notation, we can obtain an equivalent physical circuit expressed in terms of
\emph{encoded} error channels,
\begin{align}
    \label{eq:expanded_phys_circ}
    & \tikz[baseline=-1em,every node/.style={scale=0.9}, scale=0.9]{
        \draw[pwire] (-12.5, 0.25) -- (3.5, 0.25);
        \draw[pwire] (-12.5, -0.25) -- (3.5, -0.25);
        \draw[pwire] (-12.5, -0.75) -- (3.5, -0.75);
        \node[gate, physical, minimum height=4em] at (3, -0.25) {$\Lambda_P$};
        \node[gate, physical, minimum height=4em] at (2, -0.25) {$E$};
        \node[gate, physical, minimum height=4em] at (1, -0.25) {$\overline{U}_1$};
        \node[gate, physical, minimum height=4em] at (0, -0.25) {$\overline{\Lambda}$};
        \node[gate, physical, minimum height=4em] at (-1, -0.25) {$\overline{\Rec}$};
        \node[gate, physical, minimum height=4em] at (-2, -0.25) {$\overline{\Lambda}_{\Rec}$};
        \node[gate, physical, minimum height=4em] at (-3, -0.25) {$\overline{U}_2$};
        \node[gate, physical, minimum height=4em] at (-4, -0.25) {$\overline{\Lambda}$};
        \node[gate, physical, minimum height=4em] at (-5, -0.25) {$\overline{\Rec}$};
        \node[gate, physical, minimum height=4em] at (-6, -0.25) {$\overline{\Lambda}_{\Rec}$};
        \node[gate, physical, minimum height=4em] at (-7, -0.25) {$\overline{U}^{-1}$};
        \node[gate, physical, minimum height=4em] at (-8, -0.25) {$\overline{\Lambda}$};
        \node[gate, physical, minimum height=4em] at (-9, -0.25) {$\overline{\Rec}$};
        \node[gate, physical, minimum height=4em] at (-10, -0.25) {$\overline{\Lambda}_{\Rec}$};
        \node[gate, physical, minimum height=4em] at (-11, -0.25) {$D$};
        \node[gate, physical, minimum height=4em] at (-12, -0.25) {$\Lambda_M$};
        \node[] at (4, 0.25) {$\ket{\psi}$};
        \node[] at (4, -0.25) {$\ket{0}$};
        \node[] at (4, -0.75) {$\ket{0}$};
    }\ .
\end{align}
Note that above we have chosen different conventions for the encoder and decoder,
namely that $\En = \mathcal E \LP   $ and $\De = \LM \mathcal  D$.
The order in each convention is independently valid, and choosing them asymmetrically for the
encoder and decoder will make later circuit manipulations much easier.

An important concept in what follows is the mapping of physical degrees of freedom into
the logical degrees of freedom with which we are primarily interested.
In this appendix colors are used to indicate whether an operation or wire is a physical degree of freedom (\physical{represented by gray}), logical degree of freedom (\logical{represented by yellow}), or a syndrome degree of freedom (\syndrome{represented by blue}).
To isolate the effect of errors on logical degrees of freedom, it is convenient to
write the physical Hilbert space $\physical{\Hil}$ for our LRB protocol as
a tensor product  $\physical{\Hil} = \HilL \otimes \HilS$ of the logical and syndrome spaces, respectively $\HilL$
and $\HilS$, of a quantum error correcting code $\mathcal{C}$.
This representation is generic, for instance,
if $\mathcal{C}$ is a stabilizer code, in which case the computational basis of $\HilS$ will
 label the cosets of the stabilizer group. By convention, we will label the syndrome
 state corresponding to ``no error'' by the all-zero computational basis state, so that
 the encoder $E$ acts on logical states $\ket{\psi}$ as
 \begin{align}
     E (\ketl{\psi} \otimes \kets{0}) = \ket{\overline{\psi}}.
 \end{align}
 Similarly, let the decoder $D$ be the inverse of $E$, $D = E^\dagger$.
 
 For a generic quantum operation $\overline{F}$ acting on the physical (encoded) space, we use as a gadget conjugation by the perfect encoder and decoder to represent the operation in a convenient basis. Pictorially, this is represented by the circuit
\begin{align}
    & \tikz[baseline=-1\baselineskip]{
        \draw[pwire] (-6.5, 0.25) -- (-5.5, 0.25);
        \draw[pwire] (-6.5, -0.25) -- (-5.5, -0.25);
        \draw[pwire] (-6.5, -0.75) -- (-5.5, -0.75);
        \draw[pwire] (-2, 0.25) -- (-1.5, 0.25);
        \draw[pwire] (-2, -0.25) -- (-1.5, -0.25);
        \draw[pwire] (-2, -0.75) -- (-1.5, -0.75);
        \draw[lwire] (-4, 0.25) -- (-2, 0.25);
        \draw[swire] (-4, -0.25) -- (-2, -0.25);
        \draw[swire] (-4, -0.75) -- (-2, -0.75);
        \draw[pwire] (-4.5, 0.25) -- (-4, 0.25);
        \draw[pwire] (-4.5, -0.25) -- (-4, -0.25);
        \draw[pwire] (-4.5, -0.75) -- (-4, -0.75);
        \node[gate, physical, minimum height=4em] at (-6, -0.25) {$\overline{F}$};
        \node[] at (-5, -0.25) {$=$};
        \node[gate, physical, minimum height=4em] at (-2, -0.25) {$D$};
        \node[gate, physical, minimum height=4em] at (-3, -0.25) {$F$};
        \node[gate, physical, minimum height=4em] at (-4, -0.25) {$E$};
    }\ .
\end{align}
We call the conjugation by $E$ an encoding gadget. It helps us reason about the relation of the physical and logical action.

Notice we have adopted the convention that the encoder and decoder are such
that the logical degrees of freedom are on the {\color{logical} top wire} and
the syndrome degrees are on the {\color{syndrome} lower wires}, so that the
encoder and decoder act as
\begin{align}
    & \tikz[baseline=-1\baselineskip]{
        \draw[pwire] (-2, 0.25) -- (-1.5, 0.25);
        \draw[pwire] (-2, -0.25) -- (-1.5, -0.25);
        \draw[pwire] (-2, -0.75) -- (-1.5, -0.75);
        \draw[lwire] (-4.5, 0.25) -- (-2, 0.25);
        \draw[swire] (-4.5, -0.25) -- (-2, -0.25);
        \draw[swire] (-4.5, -0.75) -- (-2, -0.75);
        \draw[pwire] (-5, 0.25) -- (-4.5, 0.25);
        \draw[pwire] (-5, -0.25) -- (-4.5, -0.25);
        \draw[pwire] (-5, -0.75) -- (-4.5, -0.75);
        \node[gate, physical, minimum height=4em] at (-2, -0.25) {$D$};
        \node at (-3.25, 0.5) {\color{logical} logical};
        \node at (-3.25, -0.5) {\color{syndrome} syndrome};
        \node[gate, physical, minimum height=4em] at (-4.5, -0.25) {$E$};
    }\ .
\end{align}
With the encoding gadget, we can define several relevant examples for our protocol.
For example, conjugating by the decoder, we can write the noise channel
$\overline{\Lambda}$ and the recovery operation in terms of their effects
on the logical space as
\def\subwidth{16.5em}
    \begin{align}
    & \tikz[baseline=-1\baselineskip]{
        \begin{scope}[shift={(-2,0)}]
            \draw[pwire] (-0.5-6, 0.25) -- (0.5-6, 0.25);
            \draw[pwire] (-0.5-6, -0.25) -- (0.5-6, -0.25);
            \draw[pwire] (-0.5-6, -0.75) -- (0.5-6, -0.75);
            \node[gate, physical, minimum height=4em] at (-6, -0.25) {$\overline{\Lambda}$};
            \node[] at (-5, -0.25) {$=$};
            \draw[pwire] (-2, 0.25) -- (-1.5, 0.25);
            \draw[pwire] (-2, -0.25) -- (-1.5, -0.25);
            \draw[pwire] (-2, -0.75) -- (-1.5, -0.75);
            \draw[lwire] (-4, 0.25) -- (-2, 0.25);
            \draw[swire] (-4, -0.25) -- (-2, -0.25);
            \draw[swire] (-4, -0.75) -- (-2, -0.75);
            \draw[pwire] (-4.5, 0.25) -- (-4, 0.25);
            \draw[pwire] (-4.5, -0.25) -- (-4, -0.25);
            \draw[pwire] (-4.5, -0.75) -- (-4, -0.75);
            \node[gate, physical, minimum height=4em] at (-2, -0.25) {$D$};
            \node[gate, physical, minimum height=4em] at (-3, -0.25) {$\Lambda$};
            \node[gate, physical, minimum height=4em] at (-4, -0.25) {$E$};
        \end{scope}
        \node[] at (-2, -0.5\baselineskip) {and};
        \begin{scope}[shift={(\subwidth,0)}]
            \draw[pwire] (-0.5-6, 0.25) -- (0.5-6, 0.25);
            \draw[pwire] (-0.5-6, -0.25) -- (0.5-6, -0.25);
            \draw[pwire] (-0.5-6, -0.75) -- (0.5-6, -0.75);
            \node[gate, physical, minimum height=4em] at (-6, -0.25) {$\overline{\mathcal R}$};
            \node[] at (-5, -0.25) {$=$};
            \draw[pwire] (-2, 0.25) -- (-1.5, 0.25);
            \draw[pwire] (-2, -0.25) -- (-1.5, -0.25);
            \draw[pwire] (-2, -0.75) -- (-1.5, -0.75);
            \draw[lwire] (-4, 0.25) -- (-2, 0.25);
            \draw[swire] (-4, -0.25) -- (-2, -0.25);
            \draw[swire] (-4, -0.75) -- (-2, -0.75);
            \draw[pwire] (-4.5, 0.25) -- (-4, 0.25);
            \draw[pwire] (-4.5, -0.25) -- (-4, -0.25);
            \draw[pwire] (-4.5, -0.75) -- (-4, -0.75);
            \draw[swire]  (-3, 0) -- (-3, -0.75);
            \node[gate, physical, minimum height=4em] at (-2, -0.25) {$D$};
            \node[gate, logical, minimum height=1.5em] at (-3, 0.25) {$\mathcal R$};
            \node[gate, physical, minimum height=4em] at (-4, -0.25) {$E$};
            \draw[fill=syndrome] (-3, -0.25)  circle (0.075);
            \draw[fill=syndrome] (-3, -0.75)  circle (0.075);
        \end{scope}
    }\ .
\end{align}
Note that, as in the main text, we have only considered direct measurement of the syndromes.
In \apxref{apx:ancilla_coupled} we showed how to reduce reduce the ancilla coupled measurement to the direct case.
Following this, our derivation also holds for ancilla-coupled measurements.

The encoded gates $\{\overline{U}_i\}$ used to construct benchmarking sequences
have a nice representation 
 
\begin{subequations}
    \label{eq:flamo_the_boss_man}
    \begin{align}
        \overline{U}_i & \defeq E \bullet (\logical{U}_i \otimes \syndrome{\id}) \\               
        \tikz[baseline=-1em]{ 
            \draw[pwire] (-0.5, 0.25) -- (0.5, 0.25);
            \draw[pwire] (-0.5, -0.25) -- (0.5, -0.25);
            \draw[pwire] (-0.5, -0.75) -- (0.5, -0.75);
            \node[gate, physical, minimum height=4em] at (0, -0.25) {$\overline{U}_i$};
        }                   
        & = \tikz[baseline=-1em]{
            \draw[pwire] (-2, 0.25) -- (-1.5, 0.25);
            \draw[pwire] (-2, -0.25) -- (-1.5, -0.25);
            \draw[pwire] (-2, -0.75) -- (-1.5, -0.75);
            \draw[lwire] (-4, 0.25) -- (-2, 0.25);
            \draw[swire] (-4, -0.25) -- (-2, -0.25);
            \draw[swire] (-4, -0.75) -- (-2, -0.75);
            \draw[pwire] (-4.5, 0.25) -- (-4, 0.25);
            \draw[pwire] (-4.5, -0.25) -- (-4, -0.25);
            \draw[pwire] (-4.5, -0.75) -- (-4, -0.75);
            \node[gate, physical, minimum height=4em] at (-2, -0.25) {$D$};
            \node[gate, logical, minimum height=1.5em] at (-3, 0.25) {$U_i$};
            \node[gate, physical, minimum height=4em] at (-4, -0.25) {$E$};
        }\ .
    \end{align}
\end{subequations}
where $A \bullet B \defeq A B A^\dagger$ is the group action of the unitary group.

\begin{Remark}[\textbf{Encoder and decoder convention}]
    \label{rem:ed_conven}
    Recall for stabilizer codes, logical unitarites are only defined modulo a stabilizer operation on the encoded space.
    Thus in making the definition in \autoref{eq:flamo_the_boss_man}, we have assumed a particular convention for the encoded
    unitaries $\overline{U}_i$ and the syndrome space $\HilS$, namely that we choose representatives
    for each logical unitary that preserve our choice of basis for $\HilS$. In general, for any stabilizer
    group element $S$, $S (E \bullet (\logical{U}_i \otimes \id))$ is also a representative of the same
    logical operator, since the action of $S$ on the syndrome space is to flip the sign on each of the
    fundamental errors (destabilizers \cite{AaroGott04}). That is,
    there exists a bitstring $s(S) = b_1 \cdots b_{n - k}$ such that
    $S (E \bullet (\logical{U}_i \otimes \id)) = E \bullet (\logical{U}_i \otimes \syndrome{X}^{b_1} \otimes \cdots \otimes \syndrome{X}^{b_{n-k}})$.
    Our convention can then be seen as demanding that $s(S)$ is the all-zeros bitstring.
    Similar protocols can be derived for other such conventions.
\end{Remark}

When considering products of channels, e.g. $\overline{\Lambda} \overline{ U }_i$, some of the encoders and decoders cancel, as we have chosen the encoder and decoder such that $DE = ED = \id$. Applying this in the case of $\overline{\Lambda} \overline{ U }_i$ gives that we can express the errors
incurred by the imperfect implementation of $U_i$ in terms of their effects on the logical and syndrome registers,
\begin{subequations}
    \begin{align}
        \tikz[baseline=-1em]{
            \draw[pwire] (-1.5, 0.25) -- (0.5, 0.25);
            \draw[pwire] (-1.5, -0.25) -- (0.5, -0.25);
            \draw[pwire] (-1.5, -0.75) -- (0.5, -0.75);
            \node[gate, physical, minimum height=4em] at (0, -0.25) {$\overline{ U }_i$};
            \node[gate, physical, minimum height=4em] at (-1, -0.25) {$\overline{\Lambda}$};
        }
        & = \tikz[baseline=-1em]{
            \draw[pwire] (-3, 0.25) -- (-2.5, 0.25);
            \draw[pwire] (-3, -0.25) -- (-2.5, -0.25);
            \draw[pwire] (-3, -0.75) -- (-2.5, -0.75);
            \draw[pwire] (-6, 0.25) -- (-5, 0.25);
            \draw[pwire] (-6, -0.25) -- (-5, -0.25);
            \draw[pwire] (-6, -0.75) -- (-5, -0.75);
            \draw[lwire] (-5, 0.25) -- (-3, 0.25);
            \draw[swire] (-5, -0.25) -- (-3, -0.25);
            \draw[swire] (-5, -0.75) -- (-3, -0.75);
            \node[gate, physical, minimum height=4em] at (-3, -0.25) {$D$};
            \draw[lwire] (-8, 0.25) -- (-6, 0.25);
            \draw[swire] (-8, -0.25) -- (-6, -0.25);
            \draw[swire] (-8, -0.75) -- (-6, -0.75);
            \draw[pwire] (-8.5, 0.25) -- (-8, 0.25);
            \draw[pwire] (-8.5, -0.25) -- (-8, -0.25);
            \draw[pwire] (-8.5, -0.75) -- (-8, -0.75);
            \node[gate, logical, minimum height=1.5em] at (-4, 0.25) {$U_i$};
            \node[gate, physical, minimum height=4em] at (-5, -0.25) {$E$};
            \node[gate, physical, minimum height=4em] at (-6, -0.25) {$D$};
            \node[gate, physical, minimum height=4em] at (-7, -0.25) {$\Lambda$};
            \node[gate, physical, minimum height=4em] at (-8, -0.25) {$E$};
        } \\
        & = \tikz[baseline=-1em]{
            \draw[pwire] (-3, 0.25) -- (-2.5, 0.25);
            \draw[pwire] (-3, -0.25) -- (-2.5, -0.25);
            \draw[pwire] (-3, -0.75) -- (-2.5, -0.75);
            \draw[pwire] (-6.5, 0.25) -- (-6, 0.25);
            \draw[pwire] (-6.5, -0.25) -- (-6, -0.25);
            \draw[pwire] (-6.5, -0.75) -- (-6, -0.75);
            \draw[lwire] (-6, 0.25) -- (-3, 0.25);
            \draw[swire] (-6, -0.25) -- (-3, -0.25);
            \draw[swire] (-6, -0.75) -- (-3, -0.75);
            \node[gate, physical, minimum height=4em] at (-3, -0.25) {$D$};
            \node[gate, logical, minimum height=1.5em] at (-4, 0.25) {$U_i$};
            \node[gate, physical, minimum height=4em] at (-5, -0.25) {$\Lambda$};
            \node[gate, physical, minimum height=4em] at (-6, -0.25) {$E$};
        }\ .
    \end{align}
\end{subequations}

The situation is somewhat more complicated in expanding the definition of the noisy recovery
operator $\Recb$, as there is a significant freedom in how we define the error channels.
Using leftward triangles to represent measurements and rightward triangles to represent
preparations, we will expand $\Recb = \overline{\Lambda}_{\Rec} \overline{\Rec}$
by using a coherent implementation of the recovery procedure,
\begin{subequations}
    \label{eq:trianglenotation}
    \begin{align}
        \tikz[baseline=-1em]{
            \draw[pwire] (-1.5, 0.25) -- (0.5, 0.25);
            \draw[pwire] (-1.5, -0.25) -- (0.5, -0.25);
            \draw[pwire] (-1.5, -0.75) -- (0.5, -0.75);
            \node[gate, physical, minimum height=4em] at (0, -0.25) {$\overline{ \Rec }$};
            \node[gate, physical, minimum height=4em] at (-1, -0.25) {$\overline{\Lambda}_{ \Rec }$};
        }
        & = \tikz[baseline=-1em]{
            \draw[pwire] (-3, 0.25) -- (-2.5, 0.25);
            \draw[pwire] (-3, -0.25) -- (-2.5, -0.25);
            \draw[pwire] (-3, -0.75) -- (-2.5, -0.75);
            \draw[lwire] (-5, 0.25) -- (-3, 0.25);
            \draw[swire] (-5, -0.25) -- (-3, -0.25);
            \draw[swire] (-5, -0.75) -- (-3, -0.75);
            \draw[pwire] (-6, 0.25) -- (-5, 0.25);
            \draw[pwire] (-6, -0.25) -- (-5, -0.25);
            \draw[pwire] (-6, -0.75) -- (-5, -0.75);
            \draw[lwire] (-12, 0.25) -- (-6, 0.25);
            \draw[swire] (-9, -0.25) -- (-6, -0.25);
            \draw[swire] (-9, -0.75) -- (-6, -0.75);
            \draw[swire] (-12, -0.25) -- (-10, -0.25);
            \draw[swire] (-12, -0.75) -- (-10, -0.75);
            \draw[pwire] (-12.5, 0.25) -- (-12, 0.25);
            \draw[pwire] (-12.5, -0.25) -- (-12, -0.25);
            \draw[pwire] (-12.5, -0.75) -- (-12, -0.75);
            \draw[swire]  (-4, 0) -- (-4, -0.75);
            \node[gate, physical, minimum height=4em] at (-3, -0.25) {$D$};
            \node[gate, logical, minimum height=1.5em] at (-4, 0.25) {$\mathcal R$};
            \draw[fill=syndrome] (-4, -0.25)  circle (0.075);
            \draw[fill=syndrome] (-4, -0.75)  circle (0.075);
            \node[gate, physical, minimum height=4em] at (-5, -0.25) {$E$};
            \node[gate, physical, minimum height=4em] at (-6, -0.25) {$D$};
            \node[gate, physical, minimum height=4em] at (-7, -0.25) {$\Lambda_{\Rec}$};
            \node[gate, syndrome, minimum height=2.5em] at (-8, -0.5) {$\LM$};
            \node[meas, syndrome, scale=0.8] at (-9, -0.5) {};
            \node[prep, syndrome, scale=0.8] at (-10, -0.5) {};
            \node[gate, syndrome, minimum height=2.5em] at (-11, -0.5) {$\LP$};
            \node[gate, physical, minimum height=4em] at (-12, -0.25) {$E$};
        } \\
        & = \tikz[baseline=-1em]{
            \draw[pwire] (-3, 0.25) -- (-2.5, 0.25);
            \draw[pwire] (-3, -0.25) -- (-2.5, -0.25);
            \draw[pwire] (-3, -0.75) -- (-2.5, -0.75);
            \draw[lwire] (-9, 0.25) -- (-3, 0.25);
            \draw[swire] (-6, -0.25) -- (-3, -0.25);
            \draw[swire] (-6, -0.75) -- (-3, -0.75);
            \draw[swire] (-9, -0.25) -- (-7, -0.25);
            \draw[swire] (-9, -0.75) -- (-7, -0.75);
            \draw[pwire] (-9.5, 0.25) -- (-9, 0.25);
            \draw[pwire] (-9.5, -0.25) -- (-9, -0.25);
            \draw[pwire] (-9.5, -0.75) -- (-9, -0.75);
            \draw[swire]  (-4, 0) -- (-4, -0.75);
            \node[gate, physical, minimum height=4em] at (-3, -0.25) {$D$};
            \node[gate, logical, minimum height=1.5em] at (-4, 0.25) {$\mathcal R$};
            \draw[fill=syndrome] (-4, -0.25)  circle (0.075);
            \draw[fill=syndrome] (-4, -0.75)  circle (0.075);
            \node[gate, physical, minimum height=4em] at (-5, -0.25) {$\Lambda_{\Rec}'$};
            \node[meas, syndrome, scale=0.8] at (-6, -0.5) {};
            \node[prep, syndrome, scale=0.8] at (-7, -0.5) {};
            \node[gate, syndrome, minimum height=2.5em] at (-8, -0.5) {$\LP$};
            \node[gate, physical, minimum height=4em] at (-9, -0.25) {$E$};
        }\ .
    \end{align}
\end{subequations}
It is vital to our protocol that there is no correlations between ancilla in successive rounds of error correction. For this reason, we have included a refresh of the ancilla in the recovery operation.  The refresh is accomplished by measurement and subsequent state re-preparation. In the circuit this indicated by the triangular gates whose directionality denotes state preparation  $\blacktriangleright$ or measurement $\blacktriangleleft$.
On the second line of \autoref{eq:trianglenotation} we have composed the
syndrome measurement error with the recovery error
$\Lambda_{\Rec}'=\Lambda_{M} \Lambda_{\Rec}$.

Next, we pull the map
$\Lambda_{\Rec}'$ through the perfect recovery operation $\Rec$ to arrive at
an equivalent quantum operation $\Lambda_{\Rec}' \Rec = \Rec \Lambda_{\Rec}^*$,
where $\Lambda_{\Rec}^*$ is the dual channel of $\Lambda_{\Rec}$.
Explicitly, suppose $\Lambda_{\Rec}'$ has  Kraus operators $\{M_k\}$
and the recovery operation is the unitary $U_\Rec$.
Then $(\Lambda_{\Rec}' \mathcal U_\Rec)(\rho) = \sum_k M_k U_\Rec \rho U_\Rec\dg M_k\dg$.
Consider the
quantum operation $(\mathcal U_\Rec\Lambda_{\Rec}^*)(\rho) = \sum_k  U_\Rec
A_k\rho  A_k\dg U_\Rec\dg$. Then $\Lambda_{\Rec}'  \mathcal U_\Rec= \mathcal
U_\Rec\Lambda_{\Rec}^*$ if $A_k = U_\Rec\dg M_k U_\Rec$.

We can therefore rewrite
the final equation as
\begin{align}
    \tikz[baseline=-1em]{
        \draw[pwire] (-1.5, 0.25) -- (0.5, 0.25);
        \draw[pwire] (-1.5, -0.25) -- (0.5, -0.25);
        \draw[pwire] (-1.5, -0.75) -- (0.5, -0.75);
        \node[gate, physical, minimum height=4em] at (0, -0.25) {$\overline{ \Rec }$};
        \node[gate, physical, minimum height=4em] at (-1, -0.25) {$\overline{\Lambda}_{ \Rec }$};
    }
    & = \tikz[baseline=-1em]{
        \draw[pwire] (-3, 0.25) -- (-2.5, 0.25);
        \draw[pwire] (-3, -0.25) -- (-2.5, -0.25);
        \draw[pwire] (-3, -0.75) -- (-2.5, -0.75);
        \draw[lwire] (-9, 0.25) -- (-3, 0.25);
        \draw[swire] (-6, -0.25) -- (-3, -0.25);
        \draw[swire] (-6, -0.75) -- (-3, -0.75);
        \draw[swire] (-9, -0.25) -- (-7, -0.25);
        \draw[swire] (-9, -0.75) -- (-7, -0.75);
        \draw[pwire] (-9.5, 0.25) -- (-9, 0.25);
        \draw[pwire] (-9.5, -0.25) -- (-9, -0.25);
        \draw[pwire] (-9.5, -0.75) -- (-9, -0.75);
        \node[gate, physical, minimum height=4em] at (-3, -0.25) {$D$};
        \node[gate, logical, minimum height=1.5em] at (-5, 0.25) {$\mathcal R$};
        \draw[swire]  (-5, 0) -- (-5, -0.75);
        \draw[fill=syndrome] (-5, -0.25)  circle (0.075);
        \draw[fill=syndrome] (-5, -0.75)  circle (0.075);
        \node[gate, physical, minimum height=4em] at (-4, -0.25) {$\Lambda_{\Rec}^*$};
        \node[meas, syndrome, scale=0.8] at (-6, -0.5) {};
        \node[prep, syndrome, scale=0.8] at (-7, -0.5) {};
        \node[gate, syndrome, minimum height=2.5em] at (-8, -0.5) {$\Lambda_{P}$};
        \node[gate, physical, minimum height=4em] at (-9, -0.25) {$E$};
    }\ .
\end{align}
We recognize that this circuit also represents a classically-controlled
recovery procedure under the principle of deferred measurement, such that
our derivation is generic, and does not require experimentally implementing
a coherently-controlled recovery operator.

With this in mind, we now want to identify the unit cell of our
sequence, so that we can make the appropriate reduction to the
Magesan \etal~model. To do this we need to consider the subcircuit
corresponding to two imperfect unitary operators,
\begin{align}
    &     \tikz[baseline=-1em]{
        \draw[pwire] (-4.75, 0.25) -- (2.75, 0.25);
        \draw[pwire] (-4.75, -0.25) -- (2.75, -0.25);
        \draw[pwire] (-4.75, -0.75) -- (2.75, -0.75);
        \node[gate, physical, minimum width=2.5em, minimum height=4em] at (2, -0.25) {$\overline{ U_i }$};
        \node[gate, physical, minimum width=2.5em, minimum height=4em] at (1, -0.25) {$\overline{ \Lambda }$};
        \node[gate, physical, minimum width=2.5em, minimum height=4em] at (0, -0.25) {$\overline{ \Rec }$};
        \node[gate, physical, minimum width=2.5em, minimum height=4em] at (-1, -0.25) {$\overline{\Lambda}_{ \Rec }$};
        \node[gate, physical, minimum width=2.5em, minimum height=4em] at (-2, -0.25) {$\overline{ U}_{i+1}$};
        \node[gate, physical, minimum width=2.5em, minimum height=4em] at (-3, -0.25) {$\overline{ \Lambda }$};
        \node[gate, physical, minimum width=2.5em, minimum height=4em] at (-4, -0.25) {$\overline{ \Rec }$};
    }\ .
\end{align}
Expanding this makes it clear how to identify a unit cell with
inputs and outputs on the logical register alone,
\begin{align}
    & \tikz[baseline=-1em,every node/.style={font=\relsize{-1}}]{
        \draw[pwire] (-5, 0.25) -- (-4.5, 0.25);
        \draw[pwire] (-5, -0.25) -- (-4.5, -0.25);
        \draw[pwire] (-5, -0.75) -- (-4.5, -0.75);
        \draw[lwire] (-19, 0.25) -- (-5, 0.25);
        \draw[swire] (-10, -0.25) -- (-5, -0.25);
        \draw[swire] (-10, -0.75) -- (-5, -0.75);
        \draw[swire] (-16, -0.25) -- (-11, -0.25);
        \draw[swire] (-16, -0.75) -- (-11, -0.75);
        \draw[swire] (-19, -0.25) -- (-17, -0.25);
        \draw[swire] (-19, -0.75) -- (-17, -0.75);
        \draw[pwire] (-19.5, 0.25) -- (-19, 0.25);
        \draw[pwire] (-19.5, -0.25) -- (-19, -0.25);
        \draw[pwire] (-19.5, -0.75) -- (-19, -0.75);
        \node[gate, physical, minimum height=4em] at (-5, -0.25) {$D$};
        \node[gate, logical, minimum height=1.5em] at (-6, 0.25) {$U_i$};
        \node[gate, physical, minimum height=4em] at (-7, -0.25) {$\Lambda$};
        \node[gate, physical, minimum height=4em] at (-8, -0.25) {$\Lambda_{\Rec}^*$};
        \node[gate, logical, minimum height=1.5em] at (-9, 0.25) {$\mathcal R$};
        \draw[swire]  (-9, 0) -- (-9, -0.75);
        \draw[fill=syndrome] (-9, -0.25)  circle (0.075);
        \draw[fill=syndrome] (-9, -0.75)  circle (0.075);
        \node[meas, syndrome, scale=0.8] at (-10, -0.5) {};
        \node[prep, syndrome, scale=0.8] at (-11, -0.5) {};
        \node[gate, syndrome, minimum height=2.5em] at (-12, -0.5) {$\Lambda_{P}$};
        \node[gate, logical, minimum height=1.5em] at (-12, 0.25) {$U_{i+1}$};
        \node[gate, physical, minimum height=4em] at (-13, -0.25) {$\Lambda$};
        \node[gate, physical, minimum height=4em] at (-14, -0.25) {$\Lambda_{\Rec}^*$};
        \node[gate, logical, minimum height=1.5em] at (-15, 0.25) {$\mathcal R$};
        \draw[swire]  (-15, 0) -- (-15, -0.75);
        \draw[fill=syndrome] (-15, -0.25)  circle (0.075);
        \draw[fill=syndrome] (-15, -0.75)  circle (0.075);
        \node[meas, syndrome, scale=0.8] at (-16, -0.5) {};
        \node[prep, syndrome, scale=0.8] at (-17, -0.5) {};
        \node[gate, syndrome, minimum height=2.5em] at (-18, -0.5) {$\Lambda_{P}$};
        \node[gate, physical, minimum height=4em] at (-19, -0.25) {$E$};
    }\ .
\end{align}
With the above definition, we now compose $\Lambda' = \Lambda_{\Rec}^* \Lambda $, giving us
our desired logical unit cell,
\begin{align}
    \label{eq:lambdaprime}
    & \tikz[baseline=-1em, every node/.style={font=\relsize{-1}}]{
        \draw[pwire] (-5, 0.25) -- (-4.5, 0.25);
        \draw[pwire] (-5, -0.25) -- (-4.5, -0.25);
        \draw[pwire] (-5, -0.75) -- (-4.5, -0.75);
        \draw[lwire] (-17, 0.25) -- (-5, 0.25);
        \draw[swire] (-9, -0.25) -- (-5, -0.25);
        \draw[swire] (-9, -0.75) -- (-5, -0.75);
        \draw[swire] (-14, -0.25) -- (-10, -0.25);
        \draw[swire] (-14, -0.75) -- (-10, -0.75);
        \draw[swire] (-17, -0.25) -- (-15, -0.25);
        \draw[swire] (-17, -0.75) -- (-15, -0.75);
        \draw[pwire] (-17.5, 0.25) -- (-17, 0.25);
        \draw[pwire] (-17.5, -0.25) -- (-17, -0.25);
        \draw[pwire] (-17.5, -0.75) -- (-17, -0.75);
        \node[gate, physical, minimum height=4em] at (-5, -0.25) {$D$};
        \node[gate, logical, minimum height=1.5em] at (-6, 0.25) {$U_i$};
        \node[gate, physical, minimum height=4em] at (-7, -0.25) {$\Lambda'$};
        \node[gate, logical, minimum height=1.5em] at (-8, 0.25) {$\mathcal R$};
        \draw[swire]  (-8, 0) -- (-8, -0.75);
        \draw[fill=syndrome] (-8, -0.25)  circle (0.075);
        \draw[fill=syndrome] (-8, -0.75)  circle (0.075);
        \node[meas, syndrome, scale=0.8] at (-9, -0.5) {};
        \node[prep, syndrome, scale=0.8] at (-10, -0.5) {};
        \node[gate, syndrome, minimum height=2.5em] at (-11, -0.5) {$\Lambda_{P}$};
        \node[gate, logical, minimum height=1.5em] at (-11, 0.25) {$U_{i+1}$};
        \node[gate, physical, minimum height=4em] at (-12, -0.25) {$\Lambda'$};
        \node[gate, logical, minimum height=1.5em] at (-13, 0.25) {$\mathcal R$};
        \draw[swire]  (-13, 0) -- (-13, -0.75);
        \draw[fill=syndrome] (-13, -0.25)  circle (0.075);
        \draw[fill=syndrome] (-13, -0.75)  circle (0.075);
        \node[meas, syndrome, scale=0.8] at (-14, -0.5) {};
        \node[prep, syndrome, scale=0.8] at (-15, -0.5) {};
        \node[gate, syndrome, minimum height=2.5em] at (-16, -0.5) {$\Lambda_{P}$};
        \node[gate, physical, minimum height=4em] at (-17, -0.25) {$E$};
        \node[dashed,minimum height=2cm,minimum width=5.cm,draw] at (-12,-0.25) {};
        \node[] at (-12,-1.6) {Logical Unit Cell}
    }\ .
\end{align}
Now that we have identified the logical unit cell, we can use it to expand the circuit in \autoref{eq:expanded_phys_circ},
obtaining
\begin{align}
    & \tikz[baseline=-1em]{
        \foreach \dx in {1} {
            \begin{scope}[every node/.style={scale=0.7}, scale=0.7, shift={(16 - 6* \dx , 0)}]
                \node[] at (-7, 0.25) {$\ket{\psi}$};
                \node[] at (-7, -0.25) {$\ket{0}$};
                \node[] at (-7, -0.75) {$\ket{0}$};
                \draw[lwire] (-9, 0.25) -- (-7.5, 0.25);
                \draw[swire] (-9, -0.25) -- (-7.5, -0.25);
                \draw[swire] (-9, -0.75) -- (-7.5, -0.75);
                \draw[pwire] (-10, 0.25) -- (-9, 0.25);
                \draw[pwire] (-10, -0.25) -- (-9, -0.25);
                \draw[pwire] (-10, -0.75) -- (-9, -0.75);
                \draw[lwire] (-15, 0.25) -- (-10, 0.25);
                \draw[swire] (-14, -0.25) -- (-10, -0.25);
                \draw[swire] (-14, -0.75) -- (-10, -0.75);
                \node[gate, physical, minimum height=4em] at (-8, -0.25) {$\Lambda_{\rm P}$};
                \node[gate, physical, minimum height=4em] at (-9, -0.25) {$E$};
                \node[gate, physical, minimum height=4em] at (-10, -0.25) {$D$};
                \node[gate, logical, minimum height=1.5em] at (-11, 0.25) {$U_{\dx}$};
                \node[gate, physical, minimum height=4em] at (-12, -0.25) {$\Lambda'$};
                \node[gate, logical, minimum height=1.5em] at (-13, 0.25) {$\mathcal R$};
                \draw[swire]  (-13, 0) -- (-13, -0.75);
                \draw[fill=syndrome] (-13, -0.25)  circle (0.075);
                \draw[fill=syndrome] (-13, -0.75)  circle (0.075);
                \node[meas, syndrome, scale=0.8] at (-14, -0.5) {};
            \end{scope}
        }
        \foreach \dx in {2} {
            \begin{scope}[every node/.style={scale=0.7}, scale=0.7, shift={(17 -6* \dx , 0)}]
                \draw[lwire] (-15, 0.25) -- (-9, 0.25);
                \draw[swire] (-14, -0.25) -- (-10, -0.25);
                \draw[swire] (-14, -0.75) -- (-10, -0.75);
                \node[prep, syndrome, scale=0.8] at (-10, -0.5) {};
                \node[gate, syndrome, minimum height=2.5em] at (-11, -0.5) {$\Lambda_{P}$};
                \node[gate, logical, minimum height=1.5em] at (-11, 0.25) {$U_{\dx}$};
                \node[gate, physical, minimum height=4em] at (-12, -0.25) {$\Lambda'$};
                \node[gate, logical, minimum height=1.5em] at (-13, 0.25) {$\mathcal R$};
                \draw[swire]  (-13, 0) -- (-13, -0.75);
                \draw[fill=syndrome] (-13, -0.25)  circle (0.075);
                \draw[fill=syndrome] (-13, -0.75)  circle (0.075);
                \node[meas, syndrome, scale=0.8] at (-14, -0.5) {};
            \end{scope}
        }
        \begin{scope}[every node/.style={scale=0.7}, scale=0.7, shift={(0 , 0)}]
            \draw[lwire] (-14, 0.25) -- (-9, 0.25);
            \draw[swire] (-14, -0.25) -- (-10, -0.25);
            \draw[swire] (-14, -0.75) -- (-10, -0.75);
            \draw[pwire] (-15, 0.25) -- (-14, 0.25);
            \draw[pwire] (-15, -0.25) -- (-14, -0.25);
            \draw[pwire] (-15, -0.75) -- (-14, -0.75);
            \draw[lwire] (-16.5, 0.25) -- (-15, 0.25);
            \draw[swire] (-16.5, -0.25) -- (-15, -0.25);
            \draw[swire] (-16.5, -0.75) -- (-15, -0.75);
            \node[prep, syndrome, scale=0.8] at (-10, -0.5) {};
            \node[gate, syndrome, minimum height=2.5em] at (-11, -0.5) {$\Lambda_{P}$};
            \node[gate, logical, minimum height=1.5em] at (-11, 0.25) {$U^{-1}$};
            \node[gate, physical, minimum height=4em] at (-12, -0.25) {$\Lambda'$};
            \node[gate, logical, minimum height=1.5em] at (-13, 0.25) {$\mathcal R$};
            \draw[swire]  (-13, 0) -- (-13, -0.75);
            \draw[fill=syndrome] (-13, -0.25)  circle (0.075);
            \draw[fill=syndrome] (-13, -0.75)  circle (0.075);
            \node[gate, physical, minimum height=4em] at (-14, -0.25) {$E$};
            \node[gate, physical, minimum height=4em] at (-15, -0.25) {$D$};
            \node[gate, physical, minimum height=4em] at (-16, -0.25) {$\Lambda_{\rm M}$};
            \node[] at (-17.25, 0.25) {$Q$ vs $\bar Q$};
            \node[] at (-17.25, -0.5) {Tr};
        \end{scope}
    }\ \!.
\end{align}

We next assume that $\LM$ and $\LP$ can be written as tensor products
of channels supported on the logical and syndrome registers alone,
\begin{align}
&     \tikz[baseline=-0.5ex]{
\draw[lwire] (-0.5, 0.25) -- (0.5, 0.25);
\draw[swire] (-0.5, -0.25) -- (0.5, -0.25);
\draw[swire] (-0.5, -0.75) -- (0.5, -0.75);
\draw[lwire] (-2.5, 0.25) -- (-1.5, 0.25);
\draw[swire] (-2.5, -0.25) -- (-1.5, -0.25);
\draw[swire] (-2.5, -0.75) -- (-1.5, -0.75);
\node[gate, logical, minimum height=1.5em] at (0, 0.25) {$\Lambda_{\rm M}$};
\node[gate, syndrome, minimum height=2.5em] at (0, -0.5) {$\Lambda_{\rm M}$};
\node[] at (-1, -0.25) {$=$};
\node[gate, physical, minimum height=4em] at (-2, -0.25) {$\Lambda_{\rm M}$};
}
\end{align}
and 
\begin{align}
&     \tikz[baseline=-0.5ex]{
\draw[lwire] (-0.5, 0.25) -- (0.5, 0.25);
\draw[swire] (-0.5, -0.25) -- (0.5, -0.25);
\draw[swire] (-0.5, -0.75) -- (0.5, -0.75);
\draw[lwire] (-2.5, 0.25) -- (-1.5, 0.25);
\draw[swire] (-2.5, -0.25) -- (-1.5, -0.25);
\draw[swire] (-2.5, -0.75) -- (-1.5, -0.75);
\node[gate, logical, minimum height=1.5em] at (0, 0.25) {$\Lambda_{\rm P}$};
\node[gate, syndrome, minimum height=2.5em] at (0, -0.5) {$\Lambda_{P}$};
\node[] at (-1, -0.25) {$=$};
\node[gate, physical, minimum height=4em] at (-2, -0.25) {$\Lambda_{\rm P}$};
}.
\end{align}
Importantly, this assumption can be seen as a special case of the
earlier assumption that we can decouple $\HilL$ and $\HilS$ 
by measuring the syndrome register.
In this way, the above assumptions can be made without
any further loss of generality.

Using the decompositions of $\LM$ and $\LP$
and annihilating the initial and final encoder/decoder
pairs, we can simplify our logical RB circuit to a more useful
form,
\begin{align}
    \label{eq:simplified-lrb-reduction}
    & \tikz[baseline=-1em]{
        \foreach \dx in {1} {
            \begin{scope}[every node/.style={scale=0.85}, scale=0.85, shift={(16 -6* \dx , 0)}]
                \node[] at (-9, 0.25) {$\ket{\psi}$};
                \node[prep, syndrome, scale=0.8] at (-9, -0.5) {};
                \draw[lwire] (-15, 0.25) -- (-9.5, 0.25);
                \draw[swire] (-14, -0.25) -- (-9, -0.25);
                \draw[swire] (-14, -0.75) -- (-9, -0.75);
                \node[gate, logical, minimum height=1.5em] at (-10, 0.25) {$\Lambda_{\rm P}$};
                \node[gate, syndrome, minimum height=2.5em] at (-10, -0.5) {$\Lambda_{P}$};
                \node[gate, logical, minimum height=1.5em] at (-11, 0.25) {$U_{\dx}$};
                \node[gate, physical, minimum height=4em] at (-12, -0.25) {$\Lambda'$};
                \node[gate, logical, minimum height=1.5em] at (-13, 0.25) {$\mathcal R$};
                \draw[swire]  (-13, 0) -- (-13, -0.75);
                \draw[fill=syndrome] (-13, -0.25)  circle (0.075);
                \draw[fill=syndrome] (-13, -0.75)  circle (0.075);
                \node[meas, syndrome, scale=0.8] at (-14, -0.5) {};
            \end{scope}
        }
        \foreach \dx in {2} {
            \begin{scope}[every node/.style={scale=0.85}, scale=0.85, shift={(17 -6* \dx , 0)}]
                \draw[lwire] (-15, 0.25) -- (-9, 0.25);
                \draw[swire] (-14, -0.25) -- (-10, -0.25);
                \draw[swire] (-14, -0.75) -- (-10, -0.75);
                \node[prep, syndrome, scale=0.8] at (-10, -0.5) {};
                \node[gate, syndrome, minimum height=2.5em] at (-11, -0.5) {$\Lambda_{P}$};
                \node[gate, logical, minimum height=1.5em] at (-11, 0.25) {$U_{\dx}$};
                \node[gate, physical, minimum height=4em] at (-12, -0.25) {$\Lambda'$};
                \node[gate, logical, minimum height=1.5em] at (-13, 0.25) {$\mathcal R$};
                \draw[swire]  (-13, 0) -- (-13, -0.75);
                \draw[fill=syndrome] (-13, -0.25)  circle (0.075);
                \draw[fill=syndrome] (-13, -0.75)  circle (0.075);
                \node[meas, syndrome, scale=0.8] at (-14, -0.5) {};
            \end{scope}
        }
        \begin{scope}[every node/.style={scale=0.85}, scale=0.85, shift={(0, 0)}]
            \draw[lwire] (-14.5, 0.25) -- (-9, 0.25);
            \draw[swire] (-15, -0.25) -- (-10, -0.25);
            \draw[swire] (-15, -0.75) -- (-10, -0.75);
            \node[prep, syndrome, scale=0.8] at (-10, -0.5) {};
            \node[gate, syndrome, minimum height=2.5em] at (-11, -0.5) {$\Lambda_{P}$};
            \node[gate, logical, minimum height=1.5em] at (-11, 0.25) {$U^{-1}$};
            \node[gate, physical, minimum height=4em] at (-12, -0.25) {$\Lambda'$};
            \node[gate, logical, minimum height=1.5em] at (-13, 0.25) {$\mathcal R$};
            \draw[swire]  (-13, 0) -- (-13, -0.75);
            \draw[fill=syndrome] (-13, -0.25)  circle (0.075);
            \draw[fill=syndrome] (-13, -0.75)  circle (0.075);
            \node[gate, logical, minimum height=1.5em] at (-14, 0.25) {$\Lambda_{\rm M}$};
            \node[gate, syndrome, minimum height=2.5em] at (-14, -0.5) {$\Lambda_{\rm M}$};
            \node[] at (-15.5, 0.25) {$Q$ vs $\bar Q$};
            \node[meas, syndrome, scale=0.8] at (-15, -0.5) {};
        \end{scope}
    }\ ,
\end{align}
where we have replaced the state preparation and measurement with the triangle notation introduced in \autoref{eq:trianglenotation}. Clearly we can compose 
$\Lambda'$ and $\Lambda_{P}$ and so we redefine $\Lambda'= \Lambda'\Lambda_{P}$.

Finally, we define the logical channel $\logical{\Lambda_L}$ that we will use to complete the
reduction to Magesan~\etal,
\begin{subequations}
    \begin{align}
        \logical{\Lambda_L} & \defeq \tikz[baseline=-1em]{
            \draw[lwire] (-15, 0.25) -- (-10, 0.25);
            \draw[swire] (-14, -0.25) -- (-11, -0.25);
            \draw[swire] (-14, -0.75) -- (-11, -0.75);
            \node[prep, syndrome, scale=0.8] at (-11, -0.5) {};
            \node[gate, physical, minimum height=4em] at (-12, -0.25) {$\Lambda'$};
            \node[gate, logical, minimum height=1.5em] at (-13, 0.25) {$\mathcal R$};
            \draw[swire]  (-13, 0) -- (-13, -0.75);
            \draw[fill=syndrome] (-13, -0.25)  circle (0.075);
            \draw[fill=syndrome] (-13, -0.75)  circle (0.075);
            \node[meas, syndrome, scale=0.8] at (-14, -0.5) {};
        } \\
        & \hphantom{:}= \tikz[baseline=-1em]{
            \draw[lwire] (-18, 0.25) -- (-10, 0.25);
            \draw[swire] (-17, -0.25) -- (-11, -0.25);
            \draw[swire] (-17, -0.75) -- (-11, -0.75);
            \node[prep, syndrome, scale=0.8] at (-11, -0.5) {};
            \node[gate, physical, minimum height=2.5em] at (-12, -0.5) {$\Lambda_{P}$};
            \node[gate, physical, minimum height=4em] at (-13, -0.25) {$\Lambda$};
            \node[gate, logical, minimum height=1.5em] at (-14, 0.25) {$\mathcal R$};
            \draw[swire]  (-14, 0) -- (-14, -0.75);
            \draw[fill=syndrome] (-14, -0.25)  circle (0.075);
            \draw[fill=syndrome] (-14, -0.75)  circle (0.075);
            \node[gate, physical, minimum height=4em] at (-15, -0.25) {$\Lambda_{\mathcal R}$};
            \node[gate, physical, minimum height=2.5em] at (-16, -0.5) {$\Lambda_{P}$};
            \node[meas, syndrome, scale=0.8] at (-17, -0.5) {};
        }
    \end{align}
\end{subequations}
on the second line we have expanded the logical channel into all of its ghastly glory.

The reduction is now obvious, as substituting $\logical{\Lambda_L}$ into
\autoref{eq:simplified-lrb-reduction} gives the circuit
\begin{subequations}\label{eq:reducto}
    \begin{align}
        & \tikz[baseline=0.5em]{
            \node[] at (-6, 0.25) {$\ket{\psi}$};
            \draw[lwire] (-16.5, 0.25) -- (-6.5, 0.25);
            \node[gate, logical, minimum height=1.5em] at (-7, 0.25) {$\Lambda_{\rm P}$};
            \node[gate, logical, minimum height=1.5em] at (-8, 0.25) {$U_1$};
            \node[gate, logical, minimum height=1.5em] at (-9, 0.25) {$\Lambda_L$};
            \node[gate, logical, minimum height=1.5em] at (-10, 0.25) {$U_2$};
            \node[gate, logical, minimum height=1.5em] at (-11, 0.25) {$\Lambda_L$};
            \node[gate, logical, minimum height=1.5em] at (-12, 0.25) {$U_3$};
            \node[gate, logical, minimum height=1.5em] at (-13, 0.25) {$\Lambda_L$};
            \node[gate, logical, minimum height=1.5em] at (-14, 0.25) {$U^{-1}$};
            \node[gate, logical, minimum height=1.5em] at (-15, 0.25) {$\Lambda_L$};
            \node[gate, logical, minimum height=1.5em] at (-16, 0.25) {$\Lambda_{\rm M}$};
            \node[] at (-17.5, 0.25) {$\logical{Q}$ vs $\logical{\bar Q}$};
        } \\
        = {} & \tikz[baseline=0.5em]{
            \draw[lwire] (-15.5, 0.25) -- (-7.5, 0.25);
            \node[] at (-7, 0.25) { $\logical{\rho_L}$};
            \node[gate, logical, minimum height=1.5em] at (-8, 0.25) {$U_1$};
            \node[gate, logical, minimum height=1.5em] at (-9, 0.25) {$\Lambda_L$};
            \node[gate, logical, minimum height=1.5em] at (-10, 0.25) {$U_2$};
            \node[gate, logical, minimum height=1.5em] at (-11, 0.25) {$\Lambda_L$};
            \node[gate, logical, minimum height=1.5em] at (-12, 0.25) {$U_3$};
            \node[gate, logical, minimum height=1.5em] at (-13, 0.25) {$\Lambda_L$};
            \node[gate, logical, minimum height=1.5em] at (-14, 0.25) {$U^{-1}$};
            \node[gate, logical, minimum height=1.5em] at (-15, 0.25) {$\Lambda_L$};
            \node[] at (-16.5, 0.25) {$\logical{Q_\eff}$ vs $\logical{\bar Q_\eff}$};
        },
    \end{align}
\end{subequations}
where the effective logical measurement is $\logical{Q_\eff}$ and initial logical state $\logical{\rho_L}$ are
similarly defined in analogy as
\begin{align}
     \logical{\rho_L} & \defeq \Tr_{\HilS} \left[ \LLP({\op{\logical\psi}{\logical\psi}})\right]\\
\logical{Q_\eff}& \defeq \LLM\dg[\logical{Q}].
  \end{align}
We recognize \autoref{eq:reducto} as being of the form of \autoref{eq:reduction-target-eq},
completing the reduction. We can thus substitute the previous circuit into \autoref{eq:prb-deriv}
to identify the analogues of the randomized benchmarking parameters
$p$, $A$ and $B$ from the original \citet{MageGambEmer12} model.

Concretely,
 \begin{subequations}
     \begin{align}
         \label{eq:lrb-deriv-step0}
         & \expect_{\vec{i}}\left[
          \tikz[baseline=0.5em]{
\draw[lwire] (-15.5, 0.25) -- (-11.5, 0.25);
\draw[lwire] (-10.5, 0.25) -- (-6.5, 0.25);
\node[] at (-6, 0.25) { $\logical{\rho_L}$};
\node[gate, logical, minimum height=1.5em] at (-7, 0.25) {$U_1$};
\node[gate, logical, minimum height=1.5em] at (-8, 0.25) {$\Lambda_L$};
\node[gate, logical, minimum height=1.5em] at (-9, 0.25) {$U_2$};
\node[gate, logical, minimum height=1.5em] at (-10, 0.25) {$\Lambda_L$};
\node at (-11, 0.25) {$\ldots$};
\node[gate, logical, minimum height=1.5em] at (-12, 0.25) {$U_m$};
\node[gate, logical, minimum height=1.5em] at (-13, 0.25) {$\Lambda_L$};
\node[gate, logical, minimum height=1.5em] at (-14, 0.25) {$U^{-1}$};
\node[gate, logical, minimum height=1.5em] at (-15, 0.25) {$\Lambda_L$};
\node[] at (-16.5, 0.25) {$\logical{Q_\eff}$ vs $\logical{\bar Q_\eff}$};
}        
         \right] \\   
         = {} & \expect_{\vec{i}}\left[\Tr\left(
             \logical{Q_L} \logical{\Lambda_L} \logical{U}^{-1} \cdots \logical{\Lambda_L} \logical{U}_{i_1} \logical{\rho_L}
         \right)\right] \\
         \label{eq:lrb-deriv-step2}
         = {} & \logical{A_L} \logical{p}_{\logical{L}}^m + \logical{B_L},
     \end{align}
 \end{subequations}
 where
 \begin{subequations}
     \begin{align}
         \logical{A_L} & \defeq \Tr\left(\logical{Q_\eff} \logical{\Lambda_L}\left[
             \logical{\rho_L} - \frac{\logical{\id}}{\dim \HilL}
         \right]\right), \\
         \logical{B_L} & \defeq \Tr\left(\logical{Q_\eff} \logical{\Lambda_L}\left[
             \frac{\logical{\id}}{\dim \HilL}
         \right]\right), \\
         \text{and}\ \logical{p_L} & \defeq \frac{
             (\dim \HilL) F(\logical{\Lambda_L}, \logical{\id}) - 1
         }{\dim \HilL - 1}.
         \end{align}
 \end{subequations}

 As a consequence,
our RB protocol immediately gives an estimate of the \emph{logical} fidelity
$\logical{F_L} \defeq F(\logical{\Lambda_L}, \logical{\id})$, avoiding
the need to twirl over the full physical space $\Hil$, or to draw inferences about the error
model of the full system from RB characterization of a subsystem. Finally,
and as noted in the main text, that our protocol allows for the choice of recovery to be
made in postprocessing, this reduction also implies a protocol for learning the
contribution to $\logical{F_L}$ due to the recovery operator.

\end{document}